# Three-dimensional modeling of ionized gas

## II. Spectral energy distributions of massive and very massive stars in stationary and time-dependent modeling of the ionization of metals in H II regions

J. A. Weber, A. W. A. Pauldrach, and T. L. Hoffmann

Institut für Astronomie und Astrophysik der Universität München, Scheinerstraße 1, 81679 München, Germany
e-mail: {jweber,uh10107,hoffmann}@usm.lmu.de

January 21, 2015

**ABSTRACT**

*Context.* H II regions play a crucial role in the measurement of the chemical composition of the interstellar medium and provide fundamental data about element abundances that constrain models of galactic chemical evolution. Discrepancies that still exist between observed emission line strengths and those predicted by nebular models can be partly attributed to the spectral energy distributions (SEDs) of the sources of ionizing radiation used in the models as well as simplifying assumptions made in nebular modeling.
*Aims.* One of the key influences on the nebular spectra is the metallicity, both nebular and stellar, which shows large variations even among nearby galaxies. But whereas nebular modeling often involves testing of different nebular metallicities against their influence on the predicted spectra, adequate grids of stellar atmospheres and realistic SEDs for different metallicities are still lacking. This is unfortunate because the influence of stellar metallicity on nebular line strength ratios, via its effect on the SEDs, is of similar importance as variations in the nebular metallicity. To overcome this deficiency we have computed a grid of model atmosphere SEDs for massive and very massive O-type stars covering a range of metallicities from significantly subsolar ($0.1\,Z_\odot$) to supersolar ($2\,Z_\odot$).
*Methods.* The SEDs have been computed using a state-of-the-art model atmosphere code that takes into account the attenuation of the ionizing flux by the spectral lines of all important elements and the hydrodynamics of the radiatively driven winds and their influence on the SEDs. For the assessment of the SEDs in nebular simulations we have developed a (heretofore not available) 3d radiative transfer code that includes a time-dependent treatment of the metal ionization.
*Results.* Using the SEDs in both 1d and 3d nebular models we explore the relative influence of stellar metallicity, gas metallicity, and inhomogeneity of the gas on the nebular ionization structure and emission line strengths. We find that stellar and gas metallicity are of similar importance for establishing the line strength ratios commonly used in nebular diagnostics, whereas inhomogeneity of the gas has only a subordinate influence on the global line strengths.
*Conclusions.* Nebular diagnostics as a quantitative tool for measuring the abundances in the interstellar gas can be used to its full potential only when the influence of SEDs, metallicity, and geometric structure of the nebula are taken into account. For these purposes, detailed stellar SEDs like those of our grid are an essential ingredient for the photoionization models used to predict nebular emission line spectra.

**Key words.** Radiative transfer – stars: early-type – very massive stars – stellar winds – mass-loss – H II regions

## 1. Introduction

Although in the present phase of the universe the number and mass fraction of hot massive stars is small compared to the total stellar population, the impact of these objects on their environment is decisive for the evolution of the host galaxies. For instance, metals are produced in the cores of massive stars, and in the later evolutionary stages of these objects these metals are distributed via stellar winds in the late phases and supernova explosions into the surrounding interstellar gas. This mechanism also influences the further star formation rate in these environments (cf. Woosley & Weaver 1995, François et al. 2004, Maio et al. 2010, Hirschmann et al. 2013, Sandford et al. 1982, Oey & Massey 1995, and Bisbas et al. 2011). As massive stars have short lifetimes of only a few million years, their locations and their chemical compositions are still closely correlated with the environment where they have formed. They can thus be used as tracers for the chemical states and metallicity gradients of galaxies. Furthermore, in disk galaxies hot massive stars are the most important emitters of ionizing radiation. They act as the primary energy supply sources for H II regions and they are involved in the process of energy maintenance required for the dilute, but extended, diffuse ionized gas (cf. Haffner et al. 1999, Rossa & Dettmar 2000 and Hoffmann et al. 2012). Thus, not least with respect to their luminosities of up to $L \sim 10^8\,L_\odot$ (Pauldrach et al. 2012), hot massive stars are key players in the evolution of massive star clusters, especially in starburst clusters and galaxies. The significance of these starbursts is not only that they heat the intergalactic medium and enrich it with metals, but that massive stars themselves are also useful diagnostic tools: Stellar wind lines can presently be identified in the spectra of individual O-supergiants out to distances of 20 Mpc. Due to their high star formation rates and the resulting large number of young stars, starburst galaxies show the distinctive spectral signatures of hot massive stars in the integrated spectra of starburst galaxies even at redshifts up to $z \sim 4$ (Steidel et al. 1996, Jones et al. 2013). These spectral features can provide important information about the chemical composition, the stellar populations, and thus the galactic evolution even at extragalactic distances. That this is feasible has already been demonstrated by Pettini et al. (2000), who showed that the O star wind line features can indeed be used to constrain star formation processes and the metallicity. By comparing theoretical population synthesis models with observational results, it is further possible to reconstruct the physical properties and the recent evolution of these objects (Leitherer & Heckman 1995, Leitherer et al. 1999, Leitherer et al. 2010).

We are thus close to the point of making use of complete and completely independent quantitative spectroscopic studies of the most luminous stellar objects. To realize this objective a diagnostic tool for determining the physical properties of hot stars via quantitative UV spectroscopy is required. The status of the ongoing work to construct such an advanced diagnostic tool that includes an assessment of the accuracy of the determination of the parameters involved has recently been described by Pauldrach et al. (2012). They have shown that the atmospheric models developed for massive stars are already realistic with regard to quantitative spectral UV analysis calculated along with consistent dynamics, which allows determining stellar parameters by comparing an observed UV spectrum to a set of suitable synthetic spectra. The astronomical perspectives are enormous, not only from applications of the diagnostic techniques to massive O-type stars, but also extended to the role of massive stars as tracers of the chemical composition and the population of starbursting galaxies at high redshift. In all of these applications a close connection between observations and accurate theoretical modeling of the stellar spectra is required, and this is also the case for the physical properties and observational features of H ɪɪ regions which are closely connected with the spectra of massive stars (cf. Rubin et al. 1991, Sellmaier et al. 1996, Giveon et al. 2002, Pauldrach 2003, Sternberg et al. 2003).

The analysis of H ɪɪ regions, which is primarily based on emission line diagnostics, is a powerful method to gain information about the chemical properties and evolution of galaxies. As the strength of the emission lines depends on both the properties of the gas – density, chemical composition – and the properties of the sources of ionization – luminosity, spectral energy distribution –, quantitative analyses of these lines are used to determine abundances in H ɪɪ regions in galaxies of different metallicities and to draw conclusions about the spectral energy distributions of the irradiating stellar fluxes and thus about the upper mass range of the stellar content of the clusters. (High-quality far-infrared spectra of extragalactic H ɪɪ regions taken with the Spitzer Space Telescope are discussed and interpreted, for example, by Rubin et al. 2007, Rubin et al. 2008; work is also in progress to determine abundance gradients within single galaxies, e.g., by Márquez et al. 2002, Stanghellini et al. 2010.)

One of the most important ingredients of galaxies, clusters, and structures to be considered and to be determined in this context is obviously the metallicity. As the properties of stars and nebulae are not only to a large extent influenced by the metal abundances, but these quantities are also locally produced and modified by the physical behavior of the stellar content, it is not surprising that the metallicities found in the interstellar medium differ considerably even within a local group of galaxies. The metallicity found in the Large Magellanic Cloud, for example, has a value of $Z_{LMC} \approx 0.4\,Z_\odot$, whereas the metallicity in the Small Magellanic Cloud has been determined to $Z_{SMC} \approx 0.15\,Z_\odot$ (Dufour 1984). Both values are significantly smaller than the metallicity in the Milky Way which seems to correspond roughly to the solar value. But the term "solar abundance" itself is somewhat vague, since studies in the past decade have surprisingly yielded significantly lower number fractions of the most abun-dant metals like C, N, O, and Ne (Asplund et al. 2009) than the values determined previously (e.g., Grevesse & Sauval 1998). Other examples of large differences to the galactic metallicity in the local universe are M 83 ($Z \approx 2\,Z_\odot$, Bresolin & Kennicutt 2002) and blue low surface-brightness galaxies ($Z \approx 0.1\,Z_\odot$, Roennback & Bergvall 1995). Beyond that, metallicities vary not only from galaxy to galaxy, they also vary within the disks of certain galaxies. Based on observations of H ɪɪ regions at increasing distances from the center of our Galaxy, Rudolph et al. (2006) have found a decline of the oxygen abundance, for instance.

In order to make further progress in the diagnostics of the emission line spectra of H ɪɪ regions one obviously has to account for the metallicity-dependence of the calculated spectral energy distributions (SEDs) of massive stars. One of the most important obstacles in this regard is the fact that massive stars show direct spectroscopic evidence of winds throughout their lifetime, and these winds modify the ionizing radiation of the stars considerably (cf. Pauldrach 1987) and contribute significantly to the state and energetics of the atmospheric structures in a metallicity-dependent way. Modeling hot star atmospheres is complicated by the fact that the outflow dominates the physics of the atmospheres, in particular regarding the density stratification and the radiative transfer, which are substantially modified through the presence of the macroscopic velocity field. In the frame of a consistent treatment of the hydrodynamics, the hydrodynamics influence not only the non-LTE model[1], but are in turn controlled by the line force determined by the occupation numbers and the radiative transfer of the non-LTE model (cf. Pauldrach 1987, Pauldrach et al. 1990, Pauldrach et al. 1994, Pauldrach et al. 2001, and Pauldrach et al. 2012).

One of the objectives of the present paper is to present a grid of advanced stellar wind spectra for O-type dwarfs and supergiants at different metallicities (Sect. 2) computed using hydrodynamic atmospheric models that include a full treatment of non-LTE line blocking and blanketing[2] and the radiative force. In the second part of this paper the influence of the computed SEDs on the properties of the irradiated interstellar gas is differentially investigated. In Sect. 3 we apply our computed SEDs to a series of simulations of sample H ɪɪ regions in order to investigate the dependence of the temperature and ionization structures on the ionizing spectra and the metallicity of the gas. At first we restrict the simulations to H ɪɪ regions which are illuminated by a single star and which consist of 1-dimensional (i.e., spherically symmetric) homogeneous density structures. These restrictions are dropped in the second part of this section, where the effects of multiple radiative sources and inhomogeneous density structures on H ɪɪ regions are investigated and discussed using our recently developed 3-dimensional radiative transfer models. In Sect. 4 we will finally summarize our results along with our conclusions.

---

[1] We use term non-LTE to refer to the detailed modeling of the statistical equilibrium of the occupation numbers of the atomic levels, obtained from actually solving the system of rate equations, without assuming the approximation of local thermodynamic equilibrium (LTE, where the level populations would follow a Saha-Boltzmann distribution at the local temperature and density).

[2] The effect of line blocking refers to an attenuation of the radiative flux in the EUV and UV spectral ranges due to the combined opacity of a huge number of metal lines present in hot stars in these frequency ranges. It drastically influences the ionization and excitation and the momentum transfer of the radiation field through radiative absorption and scattering processes. As a consequence of line blocking, only a small fraction of the radiation is re-emitted and scattered in the outward direction, whereas most of the energy is radiated back to the surface of the star, leading there to an increase of the temperature ("backwarming"). Due to this increase of the temperature, more radiation is emitted at lower energies, an effect known as line blanketing.

## 2. Stellar wind models for O-type dwarfs and supergiants at different metallicities

In this section we apply our method for modeling the expanding atmospheres of hot stars to a basic grid of massive O-stars. The objective of these calculations is to present ionizing fluxes and SEDs for massive dwarfs and supergiants at different metallicity which can be used for the quantitative analysis of emission line spectra of H ɪɪ regions.

### 2.1. The general concept for calculating synthetic spectra and SEDs of massive stars

Our approach to modeling the expanding atmospheres of hot, massive stars has been described in detail in a series of previous papers (Pauldrach 1987, Pauldrach et al. 1990, 1993; Pauldrach et al. 1994, Taresch et al. 1997, Haser et al. 1998, Pauldrach et al. 1998, 2001, 2004, 2012), and we just summarize the salient points here. Our method is based on the concept of homogeneous, stationary, and spherically symmetric radiation-driven atmospheres. Despite this being an approximation to some extent, it turns out to be sufficient to reproduce all important characteristics of the expanding atmospheres in quite some detail.

A complete model atmosphere calculation consists of (a) a solution of the hydrodynamics describing velocity and density of the outflow, based on radiative acceleration by Thomson, continuum, and line absorption and scattering (this is an essential aspect of the model, because the expansion of the atmosphere alters the emergent flux considerably compared to a hydrostatic atmosphere); (b) determination of the occupation numbers from a solution of the rate equations containing all important radiative and collisional processes, using sophisticated model atoms and corresponding line lists[3]; (c) calculation of the radiation field from a detailed radiative transfer solution taking into account not only continuum, but also Doppler-shifted line opacities and emissivities[4]; and (d) computation of the temperature from the requirement of radiative (absorption/emission) and thermal (heating/cooling) balance. An accelerated Lambda iteration (ALI) procedure[5] is used to achieve consistency of occupation numbers, radiative transfer, and temperature. If required, an updated radiative acceleration can be computed from the converged model, and the process repeated.

In addition, secondary effects such as the production of EUV and X-ray radiation in the cooling zones of shocks embedded in the wind and arising from the non-stationary, unstable behavior of radiation-driven winds can, together with K-shell absorption, be optionally considered (based on a parametrization of the shock jump velocity; cf. Pauldrach et al. 1994, 2001). However, these have not been included in the models presented here, since they affect only high ionization stages like O ᴠɪ which are not relevant for the analysis of emission line spectra of H ɪɪ regions.

Of course, it needs to be clarified whether the spectral energy distributions calculated by our method are realistic enough to be used in diagnostic modeling of H ɪɪ regions. Although the radiation in the ionizing spectral range cannot be directly observed, the predicted SEDs can be verified indirectly by a comparison of observed emission line strengths and those calculated by nebular models (cf. Sellmaier et al. 1996; Giveon et al. 2002; Rubin et al. 2007, 2008). A more stringent test can be provided by a comparison of the synthetic and observed UV spectra of individual massive stars, which involves hundreds of spectral signatures of various ionization stages with different ionization thresholds, and covering a large frequency range: because almost all of the ionization thresholds lie in the spectral range shortward of the hydrogen Lyman edge (cf. Pauldrach et al. 2012), and the ionization balances of all elements depend sensitively on the ionizing radiation throughout the entire wind, the ionization balance can be traced reliably through the strength and structure of the wind lines formed throughout the atmosphere. In this way a successful comparison of observed and synthetic UV spectra (Pauldrach et al. 1994, 2001, 2004, 2012) ascertains the quality of the ionization balance and thus of the SEDs.

### 2.2. Synthetic spectra and SEDs from a model grid of massive stars

Our model grid comprises massive stars with effective temperatures ranging from 30 000 to 55 000 K and luminosities from $10^5 L_\odot$ to $2.2 \cdot 10^6 L_\odot$ (Table 1). The model parameters, correspond to those used by Pauldrach et al. (2001), which were chosen in accordance with the range of values deduced from observations.

---

[3] In total 149 ionization stages of the 26 most abundant elements (H to Zn, apart from Li, Be, B, and Sc) are considered; a detailed description of the atomic models used is given in Sect. 3 and Table 1 of Pauldrach et al. 2001, and in Sect. 2 of Pauldrach et al. 1994 where several Tables and Figures illustrating and explaining the overall procedure are shown. Low-temperature dielectronic recombination is included.

[4] If different spectral lines get shifted across the same observer's-frame frequency by the velocity field in the envelope, line overlap, which is responsible for multiple-scattering events, takes place. The method used to solve this problem is an integral formulation of the transfer equation using an adaptive stepping technique on every ray (in $p, z$ geometry) in which the radiation transfer in each micro-interval is treated as a weighted sum on the microgrid,

$$I(\tau_0(p,z)) = I(\tau_n)e^{-(\tau_n - \tau_0)} + \sum_{i=0}^{n-1}\left(e^{-(\tau_i - \tau_0)}\int_{\tau_i}^{\tau_{i+1}} S(\tau)e^{-(\tau - \tau_i)}\,d\tau(p,z)\right),$$

where $I$ is the specific intensity, $S$ is the source function and $\tau$ is the optical depth. To accurately account for the variation of the line opacities and emissivities due to the Doppler shift, all line profile functions are evaluated correctly for the current microgrid-coordinates on the ray, thus effectively resolving individual line profiles (cf. Pauldrach et al. 2001); thus, the effects of line overlap and multiple scattering are naturally included. On basis of this procedure the application of the Sobolev technique gives for the radiative line acceleration

$$g_\text{lines}(r) = \frac{2\pi}{c}\frac{1}{\rho(r)}\sum_\text{lines}\chi_\text{line}(r)\int_{-1}^{+1} I_{\nu_0}(r,\mu)\frac{1 - e^{-\tau_s(r,\mu)}}{\tau_s(r,\mu)}\mu\,d\mu$$

where

$$\tau_s(r,\mu) = \chi_\text{line}(r)\frac{c}{\nu_0}\left[(1-\mu^2)\frac{v(r)}{r} + \mu^2\frac{dv(r)}{dr}\right]^{-1}$$

is the Sobolev optical depth and $\nu_0$ is the frequency at the center of each line ($\chi_\text{line}(r)$ is the local line absorption coefficient, $\mu$ is the cosine of the angle between the ray direction and the outward normal on the spherical surface element, $v(r)$ is the local velocity, and $c$ is the speed of light). A comparison of the line acceleration of strong and weak lines evaluated with the comoving-frame method and the Sobolev technique disregarding continuum interaction is presented in Fig. 5 of Pauldrach et al. (1986), and a comparison of the comoving-frame method and the Sobolev-with-continuum technique is shown in Fig. 3 of Puls & Hummer (1988), demonstrating the excellent agreement of the two methods.

[5] For the latest update of the general method see Pauldrach et al. 2014.

We compute stellar models for metallicities of 0.1, 0.4, 1.0, and $2.0 Z_\odot$ and compare the results for two different data sets for the solar metallicities, the abundances given by Grevesse & Sauval (1998) that had been used by Pauldrach et al. (2001) and the updated values published by Asplund et al. (2009). The latter values have been determined using the comparison of the solar spectrum with three-dimensional radiative transfer simulations of the solar photosphere and differ from the former determinations by up to 38% percent for the most abundant metals (see Table 2).

**The influence of the metallicity on the winds of hot stars.** As the winds of hot stars are primarily driven by metal lines, the chemical composition of the stellar atmosphere is a decisive factor controlling the strength of the winds. Although larger metallicities lead in general to stronger winds (Kudritzki et al. 1987, Pauldrach 1987, and Pauldrach et al. 2012), due to their different line strength distributions not all elements act on the winds in the same way (cf. Pauldrach 1987). For example, an increase of the abundances of the iron group elements (Fe, Ni) will increase the mass-loss rate and correspondingly decrease the terminal velocity, whereas an increase of the abundances of some lighter elements (C, N, O, S, Ne, Ar) will increase the terminal velocity and correspondingly decrease the mass-loss rate. Thus, different wind parameters may be obtained if the abundance ratios are changed, even if the total metallicity is kept constant.[6]

We will, however, not consider such second-order effects in the present work. Instead, our computed mass-loss rates shown in Table 1 are based on the values of the mass-loss rates presented by Pauldrach et al. (2001), scaled by the corresponding factors derived from the metallicity-dependence of the wind strengths exhibited by self-consistent hydrodynamical radiation-driven wind calculations (details of this procedure are described by Pauldrach et al. 2012). We have additionally applied our hydrodynamical radiation-driven wind calculations to a grid of very massive stars (VMS) with initial masses between $150 M_\odot$ and $3000 M_\odot$, effective temperatures $T_{\rm eff}$ between 40000 K and 65000 K, and metallicities $Z$ of $0.05 Z_\odot$ and $1 Z_\odot$ (Table 3; the stellar parameters are based on theoretical evolutionary models as described by Belkus et al. 2007; see also Pauldrach et al. 2012).

**Synthetic spectra and SEDs obtained from the O-star grid.** As a primary result of our computations, Figs. A.1–A.4 show the emergent ionizing fluxes together with the corresponding shapes of the continuum for the models of our grid. As can be verified from these figures the influence of the line opacities, i.e., the difference between the hypothetical continuum and the real emergent flux, increases from objects with cooler effective temperatures to those with hotter effective temperatures and from dwarfs to supergiants. This result is not surprising, since it is a consequence of the behavior of the most important dynamical parameter, the mass-loss rate $\dot{M}$, which is directly coupled to the stellar luminosity. Thus, the mass-loss rate increases with the effective temperature and the luminosity class (cf. Table 1), and the optical depths of the spectral lines, which increase accordingly, produce, along with the increasing mass-loss rate, a more pronounced line blocking effect in the wind part of the atmosphere. As is verified from Figs. A.1–A.4, this behavior saturates however for objects with effective temperatures larger than $T_{\rm eff} = 45\,000$ K, since in these cases higher main ionization stages are encountered (e. g., Fe V and Fe VI which have a smaller number of bound-bound transitions (cf. Pauldrach 1987). The effect of line blocking is thus strongest for supergiants of intermediate $T_{\rm eff}$.

---

[6] A discussion of the dependence of the mass loss rate and the corresponding consistently calculated terminal velocity on metallicity has been presented by Kudritzki et al. (1987).

**Table 1.** Mass loss rates for different metallicities derived for our grid of model stars, based on the solar abundances determined by Asplund et al. (2009).

| Model | $Z/Z_\odot$ | $R/R_\odot$ | $\log g$ (cgs) | $\log(L/L_\odot)$ | $\dot{M}$ ($10^{-6} M_\odot$/yr) |
|---|---|---|---|---|---|
| **Dwarfs** | | | | | |
| D-30 | 0.1 | 12 | 3.85 | 5.02 | 0.0019 |
| | 0.4 | | | | 0.0039 |
| | 1.0 | | | | 0.0061 |
| | 2.0 | | | | 0.0086 |
| D-35 | 0.1 | 11 | 3.80 | 5.21 | 0.014 |
| | 0.4 | | | | 0.026 |
| | 1.0 | | | | 0.042 |
| | 2.0 | | | | 0.059 |
| D-40 | 0.1 | 10 | 3.75 | 5.36 | 0.079 |
| | 0.4 | | | | 0.16 |
| | 1.0 | | | | 0.24 |
| | 2.0 | | | | 0.33 |
| D-45 | 0.1 | 12 | 3.90 | 5.72 | 0.41 |
| | 0.4 | | | | 0.83 |
| | 1.0 | | | | 1.32 |
| | 2.0 | | | | 1.9 |
| D-50 | 0.1 | 12 | 4.00 | 5.91 | 1.7 |
| | 0.4 | | | | 3.5 |
| | 1.0 | | | | 5.4 |
| | 2.0 | | | | 7.9 |
| D-55 | 0.1 | 15 | 4.10 | 6.27 | 5.2 |
| | 0.4 | | | | 12.1 |
| | 1.0 | | | | 21 |
| | 2.0 | | | | 32 |
| **Supergiants** | | | | | |
| S-30 | 0.1 | 27 | 3.00 | 5.72 | 2.3 |
| | 0.4 | | | | 3.8 |
| | 1.0 | | | | 5.2 |
| | 2.0 | | | | 6.4 |
| S-35 | 0.1 | 21 | 3.30 | 5.77 | 3.3 |
| | 0.4 | | | | 5.7 |
| | 1.0 | | | | 8.2 |
| | 2.0 | | | | 10.6 |
| S-40 | 0.1 | 19 | 3.60 | 5.92 | 2.9 |
| | 0.4 | | | | 6.4 |
| | 1.0 | | | | 10.8 |
| | 2.0 | | | | 16 |
| S-45 | 0.1 | 20 | 3.80 | 6.17 | 3.0 |
| | 0.4 | | | | 7.7 |
| | 1.0 | | | | 15 |
| | 2.0 | | | | 24 |
| S-50 | 0.1 | 20 | 3.90 | 6.35 | 6.0 |
| | 0.4 | | | | 13.1 |
| | 1.0 | | | | 21 |
| | 2.0 | | | | 38 |

The figures show that the spectral energy distributions depend not only sensitively on the stellar parameters (especially the effective temperature), but also on the metallicity.[7] This influence on the spectra is not only due to the direct line-blocking effect caused by the metals, but also indirectly due to changes in the hydrodynamic structure that occur as a consequence of the influence of the metallicity on the radiative line acceleration.

**Table 2.** Comparison of the values of the solar abundances determined by Asplund et al. (2009) (AGS) and Grevesse & Sauval (1998) (GS) for the most abundant metals. The corrections obtained by AGS are based on a realistic 3-dimensional model of the solar atmosphere (Asplund 2005) and give values which are roughly 20% to 30% lower than the previous values obtained by GS for the most abundant elements C, N, O, and Ne.

| Element | AGS | GS | correction |
|---|---|---|---|
| | $12 + \log\left(\dfrac{n}{n_H}\right)$ | $12 + \log\left(\dfrac{n}{n_H}\right)$ | (%) |
| C | $8.43 \pm 0.05$ | $8.52 \pm 0.06$ | $-19$ |
| N | $7.83 \pm 0.05$ | $7.92 \pm 0.06$ | $-19$ |
| O | $8.69 \pm 0.05$ | $8.83 \pm 0.06$ | $-28$ |
| Ne | $7.93 \pm 0.10$ | $8.08 \pm 0.06$ | $-29$ |
| Mg | $7.60 \pm 0.04$ | $7.58 \pm 0.05$ | $+5$ |
| Si | $7.51 \pm 0.03$ | $7.55 \pm 0.05$ | $-9$ |
| S | $7.12 \pm 0.03$ | $7.33 \pm 0.11$ | $-38$ |
| Ca | $6.34 \pm 0.04$ | $6.36 \pm 0.02$ | $-5$ |
| Fe | $7.50 \pm 0.04$ | $7.50 \pm 0.05$ | $0$ |
| total metallicity | $9.02$ | $9.13$ | $-22$ |

**Table 3.** Stellar parameters and mass loss rates at different metallicities for the computed models of very massive stars (cf. Pauldrach et al. 2012).

| $T_{eff}$ (K) | $Z/Z_\odot$ | $R/R_\odot$ | $\log g$ (cgs) | $\log(L/L_\odot)$ | $\dot{M}$ ($10^{-6}\ M_\odot/\mathrm{yr}$) |
|---|---|---|---|---|---|
| | | | $M_* = 150\ M_\odot$ | | |
| 40 000 | 0.05 | 40 | 3.40 | 6.57 | 4.7 |
| | 1.00 | | | | 26 |
| 50 000 | 0.05 | 25.8 | 3.79 | | 5.0 |
| | 1.00 | | | | 23 |
| | | | $M_* = 300\ M_\odot$ | | |
| 40 000 | 0.05 | 68 | 3.24 | 7.03 | 59 |
| | 1.00 | | | | 220 |
| | | | $M_* = 600\ M_\odot$ | | |
| 50 000 | 0.05 | 43.5 | 3.94 | 7.03 | 6.0 |
| | 1.00 | | | | 34 |
| | | | $M_* = 1000\ M_\odot$ | | |
| 45 000 | 0.05 | 92 | 3.51 | 7.50 | 53 |
| | 1.00 | | | | 190 |
| 65 000 | 0.05 | 44.5 | 4.14 | 7.50 | 35 |
| | 1.00 | | | | 98 |
| | | | $M_* = 3000\ M_\odot$ | | |
| 45 000 | 0.05 | 164 | 3.49 | 8.00 | 285 |
| | 1.00 | | | | 923 |
| 65 000 | 0.05 | 79 | 4.12 | 8.00 | 132 |
| | 1.00 | | | | 385 |

In Table 4 we list for each model the number of photons emitted per second capable of ionizing H, He, $He^+$, $O^+$, $Ne^+$ and $S^+$. The hydrogen ionizing flux determines (along with the density structure and temperature of the gas) the extent of the ionized volume while helium is important as absorber for the hard ionizing radiation. The ionization products of considered ionization stages of metals are effective line radiation emitters in gaseous nebulae and therefore play an important role for the corresponding line diagnostics. Their line emission also significantly contributes the energy balance of the ionized gas. The corresponding ionizing fluxes of stellar sources are therefore decisive for the properties of the gas in their environment.

Although the number of hydrogen-ionizing photons depends only weakly on metallicity in the range stellar temperature range of our grid, significant differences of up to several orders of magnitude are found for the ionization stages of $He^+$ and $Ne^+$. These pronounced anti-correlations between the ionizing fluxes and the stellar metallicity indicate strongly that the influence of stellar metallicity on nebular line strength ratios is of similar importance as that of variations in the nebular metallicity.[8] But we also note that the relative influence of the stellar metallicity on the number of emitted ionizing photons decreases for larger effective temperatures and hence harder ionizing spectra.

**Very massive stars.** As a first indication to what extent very massive stars may be identified by their ionizing influence on the environment, we list in Table 5 the photon emission rates of our very massive star models in the different ionization continua. The SEDs of the hottest of these stars with an effective temperature of 65 000 K and masses of 1000 $M_\odot$ and 3000 $M_\odot$ are characterized by ratios of He II-ionizing to H-ionizing fluxes that exceed those of even the hottest "normal" O-type model stars (the 55 000 K dwarfs) by a factor of 5 to 10, while for the cooler very massive star models the ratios are similar to those of the normal O-star models.[9] We will simulate the effects of the very massive star model SEDs on their surrounding H II regions in Sect. 3.2.2. Plots of the SEDs of the VMS are shown in Figs. A.5 and A.6 (The corresponding data sets can by copied from the links provided there).

## 3. Application of the calculated SEDs to one- and three-dimensional simulations of H II regions: influence on the emission line intensities

The ionized gas of H II regions re-emits most of the EUV energy output of the star in a limited number of lines in the UV, the optical, and the IR. The comparatively high intensities in these lines allows observing the ionized gas even if the central source

---

**Table 4.** Numerical values of the integrals of ionizing photons emitted per second for the ionization stages of H, He, He$^+$, O$^+$, Ne$^+$, and S$^+$, as well as the luminosity at the reference wavelength $\lambda = 5480$ Å. With respect to their ionization energies these ions are important in the context of emission line diagnostics. The integrals are defined as $Q_X = \int_{\nu_X}^{\infty} (L_\nu/h\nu)\,d\nu$, where $h\nu_X$ is the ionization energy of ion $X$. (The ionization energy is 1.72 Ryd for S II, 1.81 Ryd for He I, 2.77 Ryd for O II, and 3.01 Ryd for Ne II.)

| Model | $Z/Z_\odot$ | log $Q_H$ | log $Q_{He}$ | log $Q_{He^+}$ | log $Q_{O^+}$ | log $Q_{Ne^+}$ | log $Q_{S^+}$ | $L_\nu$(5480 Å) ($10^{22}$ erg/s/Hz) |
|---|---|---|---|---|---|---|---|---|
| | | | | Dwarfs | | | | |
| D-30 | 0.1 | 47.60 | 45.02 | 38.97 | 42.92 | 41.10 | 45.95 | 3.85 |
| | 0.4 | 47.58 | 45.03 | 36.49 | 42.89 | 41.12 | 45.82 | 4.09 |
| | 1.0 | 47.56 | 45.09 | 35.23 | 42.96 | 41.31 | 45.76 | 4.17 |
| | 2.0 | 47.71 | 45.34 | 35.62 | 43.26 | 41.62 | 45.86 | 4.06 |
| D-35 | 0.1 | 48.67 | 47.58 | 42.15 | 46.55 | 45.50 | 47.74 | 4.22 |
| | 0.4 | 48.66 | 47.56 | 41.51 | 46.29 | 44.53 | 47.71 | 4.21 |
| | 1.0 | 48.68 | 47.58 | 37.41 | 46.17 | 44.22 | 47.70 | 4.36 |
| | 2.0 | 48.68 | 47.49 | 37.61 | 45.83 | 44.00 | 47.63 | 4.45 |
| D-40 | 0.1 | 49.09 | 48.43 | 43.95 | 47.76 | 46.97 | 48.51 | 4.19 |
| | 0.4 | 49.08 | 48.35 | 43.67 | 47.47 | 46.41 | 48.44 | 4.30 |
| | 1.0 | 49.08 | 48.31 | 43.78 | 47.30 | 46.17 | 48.41 | 4.38 |
| | 2.0 | 49.06 | 48.25 | 43.87 | 47.07 | 45.98 | 48.36 | 4.53 |
| D-45 | 0.1 | 49.53 | 48.96 | 45.22 | 48.37 | 47.70 | 49.02 | 6.60 |
| | 0.4 | 49.52 | 48.89 | 45.24 | 48.19 | 47.43 | 48.97 | 7.27 |
| | 1.0 | 49.50 | 48.83 | 45.12 | 47.97 | 47.12 | 48.92 | 7.43 |
| | 2.0 | 49.50 | 48.78 | 45.00 | 47.77 | 46.84 | 48.88 | 7.75 |
| D-50 | 0.1 | 49.76 | 49.26 | 46.07 | 48.72 | 48.11 | 49.32 | 7.69 |
| | 0.4 | 49.75 | 49.22 | 46.17 | 48.60 | 47.91 | 49.29 | 8.44 |
| | 1.0 | 49.74 | 49.19 | 46.04 | 48.42 | 47.64 | 49.26 | 8.79 |
| | 2.0 | 49.74 | 49.16 | 45.77 | 48.28 | 47.41 | 49.23 | 9.50 |
| D-55 | 0.1 | 50.14 | 49.71 | 46.88 | 49.23 | 48.70 | 49.76 | 12.93 |
| | 0.4 | 50.14 | 49.72 | 47.18 | 49.19 | 48.54 | 49.77 | 13.62 |
| | 1.0 | 50.14 | 49.72 | 47.11 | 49.11 | 48.36 | 49.77 | 14.07 |
| | 2.0 | 50.14 | 49.69 | 46.76 | 48.90 | 48.11 | 49.74 | 16.04 |
| | | | | Supergiants | | | | |
| S-30 | 0.1 | 49.27 | 48.00 | 37.72 | 46.55 | 44.90 | 48.16 | 22.22 |
| | 0.4 | 49.25 | 47.88 | 36.04 | 46.26 | 44.06 | 48.04 | 22.97 |
| | 1.0 | 49.25 | 47.85 | 35.32 | 45.54 | 43.35 | 48.01 | 22.98 |
| | 2.0 | 49.22 | 47.78 | 35.29 | 45.18 | 41.85 | 47.94 | 23.90 |
| S-35 | 0.1 | 49.48 | 48.63 | 37.34 | 47.73 | 46.82 | 48.73 | 15.54 |
| | 0.4 | 49.47 | 48.55 | 37.45 | 47.25 | 46.18 | 48.66 | 16.27 |
| | 1.0 | 49.47 | 48.47 | 37.50 | 46.83 | 45.92 | 48.59 | 16.28 |
| | 2.0 | 49.47 | 48.36 | 36.56 | 46.21 | 45.55 | 48.49 | 16.98 |
| S-40 | 0.1 | 49.67 | 49.01 | 45.06 | 48.32 | 47.51 | 49.10 | 15.01 |
| | 0.4 | 49.67 | 48.93 | 38.94 | 48.03 | 47.11 | 49.03 | 15.70 |
| | 1.0 | 49.66 | 48.81 | 38.54 | 47.57 | 46.70 | 48.95 | 16.68 |
| | 2.0 | 49.66 | 48.70 | 38.06 | 47.07 | 46.27 | 48.86 | 18.15 |
| S-45 | 0.1 | 49.98 | 49.41 | 45.97 | 48.81 | 48.12 | 49.47 | 19.12 |
| | 0.4 | 49.97 | 49.35 | 45.79 | 48.67 | 47.78 | 49.43 | 19.62 |
| | 1.0 | 49.97 | 49.28 | 39.23 | 48.30 | 47.37 | 49.36 | 22.52 |
| | 2.0 | 49.98 | 49.19 | 38.85 | 47.92 | 46.97 | 49.29 | 23.92 |
| S-50 | 0.1 | 50.21 | 49.73 | 46.76 | 49.21 | 48.59 | 49.78 | 20.66 |
| | 0.4 | 50.20 | 49.70 | 46.79 | 49.08 | 48.34 | 49.76 | 21.97 |
| | 1.0 | 50.20 | 49.69 | 46.41 | 48.88 | 48.06 | 49.75 | 23.66 |
| | 2.0 | 50.21 | 49.69 | 46.23 | 48.84 | 47.81 | 49.75 | 24.80 |

is visually much fainter, making it possible to determine the temperatures and metallicities of the ionized gas even of extragalactic star-forming regions (cf. Zaritsky et al. 1994, Moy et al. 2001, Rubin et al. 2008, Rubin et al. 2010, Rubin et al. 2012, Pilyugin et al. 2013). Additionally, the emission lines of H II regions can be measured using narrow-band filters allowing for observations which simultaneously provide information about the emission spectrum and the spatial structure of an object. This method has been applied to examine the metallicities of H II regions as a function of the position within their host galaxies (cf. Cedrés et al. 2012) or to analyze the substructure of the gas in a single H II region (cf. Heydari-Malayeri et al. 2001).

**Table 5.** Numerical values of the integrals of ionizing photons emitted per second by very massive stars. The structure of the table corresponds to Table 4.

| $T_{\text{eff}}$ | $Z/Z_\odot$ | $\log Q_{\text{H}}$ | $\log Q_{\text{He}}$ | $\log Q_{\text{He}^+}$ | $\log Q_{\text{O}^+}$ | $\log Q_{\text{Ne}^+}$ | $\log Q_{\text{S}^+}$ | $L_\nu(5480\,\text{Å})$ $(10^{22}\,\text{erg/s/Hz})$ |
|---|---|---|---|---|---|---|---|---|
| | | | | $M_* = 150\,M_\odot$ | | | | |
| 40 000 | 0.05 | 50.32 | 49.68 | 45.87 | 49.06 | 48.29 | 49.76 | 64 |
| | 1 | 50.31 | 49.54 | 39.10 | 48.34 | 47.42 | 49.67 | 69 |
| 50 000 | 0.05 | 50.42 | 49.95 | 46.83 | 49.46 | 48.90 | 50.00 | 33 |
| | 1 | 50.42 | 49.91 | 46.85 | 49.20 | 48.33 | 49.98 | 38 |
| | | | | $M_* = 300\,M_\odot$ | | | | |
| 40 000 | 0.05 | 50.91 | 50.36 | 44.21 | 49.73 | 49.08 | 50.42 | 160 |
| | 1 | 50.97 | 50.24 | 39.25 | 48.83 | 47.62 | 50.36 | 205 |
| | | | | $M_* = 600\,M_\odot$ | | | | |
| 50 000 | 0.05 | 50.85 | 50.39 | 47.04 | 49.91 | 49.38 | 50.44 | 93 |
| | 1 | 50.85 | 50.25 | 47.36 | 49.68 | 48.99 | 50.34 | 107 |
| | | | | $M_* = 1000\,M_\odot$ | | | | |
| 45 000 | 0.05 | 51.32 | 50.60 | 46.96 | 49.76 | 48.86 | 50.69 | 398 |
| | 1 | 51.30 | 50.70 | 47.06 | 49.79 | 48.75 | 50.80 | 427 |
| 65 000 | 0.05 | 51.39 | 51.01 | 49.06 | 50.55 | 50.06 | 51.08 | 144 |
| | 1 | 51.38 | 51.03 | 49.19 | 50.64 | 50.20 | 51.08 | 153 |
| | | | | $M_* = 3000\,M_\odot$ | | | | |
| 45 000 | 0.05 | 51.84 | 51.25 | 47.99 | 50.53 | 49.74 | 51.32 | 1092 |
| | 1 | 51.83 | 50.91 | 40.68 | 49.34 | 48.44 | 51.08 | 1249 |
| 65 000 | 0.05 | 51.88 | 51.55 | 49.77 | 51.15 | 50.68 | 51.59 | 403 |
| | 1 | 51.92 | 51.50 | 49.49 | 51.04 | 50.56 | 51.55 | 378 |

In this section we investigate the influence of the computed stellar SEDs on the properties of H II regions with homogeneous and with inhomogeneous density structures. Via simulated sample H II regions we examine how the temperature and ionization structures in the gas depend on the ionizing spectra and the stellar and nebular metallicity. The question remains, however, to what extent these models may represent real-world H II regions. The metallicity, the complex geometric structures of H II regions, and the clumpiness of the medium are usually quantitatively unknown, but may affect the analysis. If discrepancies are encountered, it is often difficult to decide which of the assumptions is responsible for the disagreement found. The computed stellar SEDs themselves can be tested, if the atmospheric models are sufficiently consistent, by their own predicted UV spectra, since the spectral lines and the corresponding ionization fractions in the stellar winds are influenced by the EUV radiation field via the same atomic processes (though at different temperatures and densities) as those that shape the nebular spectra under influence of the emergent stellar flux. In a series of papers we have shown that our consistently calculated synthetic spectra are in sufficiently good agreement with the observed UV spectra of O stars (cf. Pauldrach et al. 1994, Pauldrach et al. 2001, Pauldrach et al. 2004, Pauldrach et al. 2012).

In this work we will therefore study the influence of these effects separately. We begin in Sect. 3.1 with simple homogeneous models with spherical geometry to explore the influence of the stellar SEDs, covering the range of effective temperatures and metallicities given by the grid of stellar models presented in Sect. 2.2.

Next, we will analyze how the assumption of "perfect" spherically symmetric nebulae influences the emission line diagnostics by comparing the results from homogeneous, spherical models to those from 3-dimensional models using a fractally inhomogeneous density structure (such as described by Elmegreen & Falgarone 1996 and Wood et al. 2005). The simulations of these structures will be performed using the 3d radiative transfer code described by Weber et al. (2013). This code has been extended for the present paper such that it can account for the ionization structure and the line emission of the most abundant metal ions and the resulting influence of these ions on the energy balance of the H II regions – for the first time with full time-dependence (see Sect. 3.2.1).

Third, we will compare the gas temperatures and the related line emission of evolving H II regions with the respective values of steady-state regions. The motivation is that the description of H II regions as steady-state systems is an approximation which neglects the evolution of star-forming regions (cf. Preibisch & Zinnecker 2007 and Murray 2011) and the short lifetimes of the massive stars (cf. Langer et al. 1994) that act as sources of ionization.

Real-world star-forming regions such as the Orion nebula (cf. Muench et al. 2008) and the $\eta$ Carinae region (cf. Smith 2006) contain not just a single hot star as ionizing source, but several. In the last step of this work we will therefore replace the single-star ionizing sources in the models by clusters of stars. We compute the ionization structure and total emergent fluxes of a of a single H II region illuminated by a dense cluster and compare the results with those from a simulation where larger distances between the ionizing sources lead to the formation of partly separate H II regions. Additionally we will investigate the possibility of finding very massive stars (VMS) in star-forming regions by means of the line emission from the gas ionized by the associations.

As a key ingredient in the simulations we focus here on the accurate description of the time-dependent ionization structure of the metals. Schmidt-Voigt & Koeppen (1987) had presented a computationally efficient approach where the ionization structure of a given element is determined by interpolating between the initial conditions and the stationary case using a single eigenvalue of the system of rate equations (see below), but this approach is not accurate if more than two consecutive ionization stages of an element have to be considered, because the interpolation does not account for the fact that the transition between two non-contiguous ionization stages requires the creation of ions of the intermediate stages. In the method by Graziani et al. (2013) the radiative transfer and the computation of the occupation numbers are performed consistently for hydrogen and helium. For the occupation numbers of the metal ions and the temperature of the gas pre-computed results from the Cloudy code (Ferland et al. 2013), which describes the stationary states of H II regions using spherical symmetry, are used. These results are selected from a database such that they match the 3d results for the occupation numbers of H and He ions, and the radiation field.

Our approach on the other hand is to apply a numerical method which treats the ionization stages of all the metals we account for in the same consistent way as hydrogen and helium. The occupation numbers $n \equiv n(\mathbf{r}, t)$ of all ionization stages $i, j$ of the elements considered are calculated using the equation of the time-dependent statistical "equilibrium" (Pauli Master Equation, Pauli 1928)

$$\frac{\mathrm{d}}{\mathrm{d}t} n_i(t) = \sum_{i \neq j} \mathcal{P}_{j,i} n_j(t) - \sum_{i \neq j} \mathcal{P}_{i,j} n_i(t), \tag{1}$$

which describes the temporal change of the number density of all ionization stages $i$, and contains in the rate coefficients $\mathcal{P}_{i,j}$ all important radiative ($\mathcal{R}_{i,j}$) and collisional ($C_{i,j}$) transition rates.

For the description of these systems of differential equations we chose an approach that combines integrating the condition of particle conservation within the rate matrices (as described by Mihalas 1978) with providing a robust solution for temporal evolution of the system. To realize this aim we define, following the notation by Marten (1993), a vector $\mathbf{x}$, which contains in its components the number fractions of all considered $N$ ionization stages relative to the total number density of the element and a matrix $\mathbf{G}'$, where the components are defined as $g'_{i,i} = (\sum_{j=1, j \neq k}^{N} \mathcal{P}_{i,j}) + \mathcal{P}_{i,k}$ and $g'_{i,j \neq i} = -\mathcal{P}_{j,i} + \mathcal{P}_{k,i}$. The fraction $x_k(t)$ of the ionization stage $k$ is replaced by the condition of particle conservation:

$$x_k(t) = 1 - \sum_{j=1, j \neq k}^{N} x_j(t). \tag{2}$$

The resulting inhomogeneous system of differential equations is

$$\mathbf{E}' \cdot \frac{\mathrm{d}}{\mathrm{d}t} \mathbf{x}(t) + \mathbf{G}' \cdot \mathbf{x}(t) = \mathbf{b}, \tag{3}$$

where the components of $\mathbf{b}$ are $b_i = \mathcal{P}_{k,i}$ and where all coefficients of the redundant $k$-th row of $\mathbf{G}'$ and $\mathbf{b}$ have been replaced by 1, and those of the $k$-th column of $\mathbf{E}$ by 0 (with this replacement the unity matrix $\mathbf{E}$ becomes $\mathbf{E}'$) – with these numbers inserted the corresponding components represent in total the condition of particle conservation.

For the spherically symmetric models of Sect. 3.1 we focus on the stationary case (where the time-derivative in Eq. 3 disappears) in order to make the results for the H II regions models using our new stellar SEDs as ionizing sources comparable to the results of other simulations. In Sect. 3.2 we present our results for the time-dependent ionization structure of metals in H II regions in the context of the description of our 3d radiative transfer code.

### 3.1. Spherically symmetric models of H II regions

The standard procedure for simulations of emission line spectra of H II regions is still mostly founded on spherically symmetric models (cf. Stasińska & Leitherer 1996, Hoffmann et al. 2012, and Ferland et al. 2013). They remain useful as a comparative tool because the wavelength region of the ionizing sources blueward of the Lyman edge cannot be observed directly, but the spectral energy distribution of the ionizing flux can be nevertheless studied with such models by means of their influence on the emission line spectra of gaseous nebulae. Such tests are however not altogether without uncertainties due to the additional dependence of the emission line strengths on the chemical composition of the gas in the H II regions.

Below we will outline our numerical approach to investigate these dependencies quantitatively for spherically symmetric model H II regions. The results obtained with this method will then be discussed for a grid of models with different metallicities, using stellar SEDs computed for different temperatures and metal abundances (Sect. 2.2).

#### 3.1.1. The numerical approach applied for the computation of the spherical H II region models

The basic equations describing the temperature and ionization structure of H II regions are similar to those used to describe non-LTE stellar atmospheres in statistical equilibrium. For the computation of the spherically symmetric models of gaseous nebulae we therefore use a modified version of the WM-basic stellar atmosphere code (cf. Sect. 2.1), which has been adapted to the dilute radiation fields and low gas densities of H II regions (cf. Hoffmann et al. 2012). This approach yields descriptions of steady-state H II regions, in which ionization and recombination, as well as heating and cooling are in equilibrium at every radius point. In such a stationary state Eq. 3 simplifies to

$$\mathbf{G}' \cdot \mathbf{x}^{\infty} = \mathbf{b}. \tag{4}$$

This equation must be be solved iteratively until it converges to the final value for the stationary state $\mathbf{x}^{\infty}$, because the rate coefficients which define the entries of $\mathbf{G}'$ themselves depend on the occupation numbers $\mathbf{x}$: on the one hand, the recombination and collisional ionization rate coefficients in a gas are proportional to the electron density, which in turn mainly depends on the ionization structure of the most abundant elements hydrogen and helium; on the other hand, the radiative ionization rates are determined by the mean intensity $J_\nu$, which is influenced by the ionization-dependent opacity of the matter between the radiation sources and the considered point in the simulation volume, and by the emissivity of the surrounding gas.

In our approach $J_\nu$ is computed using a radiative transfer procedure which is performed along a number of parallel rays intersecting radius shells around the source at different angles, describing the radius- and direction-dependent intensities $I_\nu$. The mean intensity at a given radius is computed by evaluating the integral $J_\nu = \frac{1}{2} \int_{-1}^{1} I_\nu \, \mathrm{d}\mu$, where $\mu$ is the cosine of the angle between a ray and the outward normal at a given radius. The diffuse radiation field created by recombination processes and electron-scattering is treated correctly, avoiding approximations regarding the propagation of photons such as "case B" or "outward-only".

Apart from the ionization structure, the emission spectrum of an H II region primarily depends on the temperature of the gas as the recombination and collisional excitation rates are functions of the temperature. The interpretation of observations of H II regions thus requires on an accurate understanding of the microphysical processes that lead to gains and losses of the thermal energy of the gas, which in turn determines the temperature structure. The processes regarded for the computation of the energy balance in H II regions are heating by photoionization and cooling by radiative recombination, as well as free-free and collisional bound-bound processes.

The low density of the interstellar gas (compared to the gas in stellar atmospheres) leads to small collisional de-excitation rates. Thus radiative transitions of collisionally excited lines are important or even the dominant cooling processes in H II regions.[10] We extend the modeling described by Hoffmann et al. (2012) to include the cooling rates connected to the forbidden radiative de-excitation processes of collisionally excited sub-states of the ground levels of C II, N II, N III, O II, O IV, Ne III, S III and S IV. The cooling by fine-structure transitions is computed by multiplying the collisional transition rates into the excited states with the probability of the corresponding radiative de-excitation processes and the energy of the photons emitted during the relaxation back into the ground state.[11]

### 3.1.2. Dependence of the properties of H II regions on the ionizing sources

In our systematic series of simulations we examine the ionization and temperature structures and the resulting emission spectra of the gas in homogeneous H II regions irradiated by single stars. In this series we consider the temperature range of O stars (30 000 K to 55 000 K) and we use the same composition for both nebular and stellar matter[12], covering the metallicity range

[10] A collisional excitation processes which is followed by the corresponding collisional de-excitation process in total does not modify the thermal energy content of the gas. The probability of a radiative de-excitation is computed as $p_{rad} = n_{crit}/(n_e + n_{crit})$, where $n_e$ is the electron density and $n_{crit}$ is the critical density, defined as the electron density for which the collisional de-excitation rate is equal to the radiative de-excitation rate. For instance, the critical density of the $^3P_1$ state of O III, which can be de-excited by the emission of an far-infrared photon with $\lambda = 88\,\mu m$, is $5.1 \cdot 10^5\,cm^{-3}$. This value is considerably lower than the critical density of $6.8 \cdot 10^5\,cm^{-3}$ for the $^1D_2$ state of the O III ion, which is de-excited by the emission of the O III 5007 Å + 4959 Å lines. We use the values from Osterbrock & Ferland 2006.

[11] The collision strengths used for the computation of the collisional excitation rates are taken from Blum & Pradhan (1992), Butler & Zeippen (1994), Lennon & Burke (1994), Tayal & Gupta (1999) (as collected in Osterbrock & Ferland 2006). The nebular approximation is applied for the computation of transition rates into collisionally excited fine-structure levels, i.e., the computation of the fine structure transitions is based on the assumption that almost all ions are in their ground state and excitations from non-ground levels do not have to be considered.

[12] The assumption that the chemical composition of H II regions is equal to the composition of the embedded stars is not necessarily realistic because the formation of dust can lead to a depletion of metals from the gas phase. Shields & Kennicutt (1995) find, for instance, a depletion of ≈ 50% for carbon, ≈ 40% for oxygen, and ≈ 20% for nitrogen for solar metallicity, while according to their results sulfur remains almost undepleted. By contrast, a significant depletion of sulfur is found by

of star-forming regions in the present-day universe. In a second series, we keep the gas at solar metallicity in order to analyze the influence of the metallicity-dependent stellar SEDs (using our 40 000 K dwarf stars) on the temperature and ionization structure of the H II regions independently from the effects of the metallicity of the gas of the H II regions.

**The dependence of the ionization structure of H II regions on the metallicity and the stellar SEDs.** In a steady-state H II region the number of recombinations the ionized volume (the Strömgren sphere) must equal the stellar emission rate of ionizing photons. Thus, the sizes of the Strömgren spheres depend on the SEDs of the ionizing stars and on the recombination rates of the ions in the H II regions.

In the presented grid the radii $r_{H II}$ of the hydrogen Strömgren spheres grow significantly for lower metallicities but otherwise equal stellar parameters $R$, $\log(g)$ and $T_{eff}$ (see Figs. A.7 and A.8). For instance, the hydrogen Strömgren radius for a metallicity of $0.1\,Z_\odot$ is approximately 50% larger than the Strömgren radius for $2.0\,Z_\odot$ in the case of the 40 000 K dwarf stars. As shown in Table 4 the hydrogen-ionizing fluxes of O stars are – for otherwise equal stellar parameters – almost independent of metallicity. The different Strömgren radii are therefore primarily a consequence of the different recombination rates, which increase for lower temperatures of the ionized gas as they occur for higher metallicities. This relation results from the fact that radiative decays of collisionally excited states of metal ions are the dominant cooling processes in the ionized gas.[13] Cooler temperatures in turn increase the recombination coefficients of H II and other ions (cf. Osterbrock & Ferland 2006) and hence reduce the ionized volumes.

The size of the He II Strömgren sphere is of particular importance because it marks it marks the boundary where all photons with energies above the He I ionization threshold have been used up, and consequently no significant amounts of metal ions requiring ionization energies above that of He I will be found. Among these ions are N III, O III, Ne III and S IV. The volumes where helium is ionized are considerably smaller than the hydrogen Strömgren spheres for stellar effective temperatures of $T_{eff} = 30\,000$ K. The radii of the hydrogen Strömgren spheres exceed the radii of the helium Strömgren spheres by a factor of approximately 3 in the case of the dwarf stars. By contrast, for the supergiants the corresponding factor is approximately 1.5. This results from a lower ratio of He-ionizing to H-ionizing photons for the dwarf models compared to the supergiant models (cf. Table 4). Helium is singly ionized up to the Strömgren radius of hydrogen in H II regions where the effective temperature of the stellar sources is ≥ 40 000 K.

The radius range in which helium is fully ionized, i.e., where He III is the most abundant ionization stage of helium, is small in comparison to the hydrogen Strömgren sphere for dwarf and supergiant O stars. These He III regions are too small to be resolved

Rubin et al. (2007, 2008). There are variations in the metal depletion rates among different H II regions as the formation and destruction of dust depends on the chemical composition and temperature of the gas, and on the radiation field.

[13] We note that in the outer regions of the H II regions with 0.1 $Z_\odot$ the collisional excitation of neutral hydrogen atoms becomes the most important important cooling process. The collisional cooling by hydrogen increases near the Strömgren radius due to the larger abundances of neutral hydrogen. The result is that there is no increase of the temperatures in the outer parts of the ionized regions, unlike for the higher metallicities, where such maxima are caused by radiation hardening.

in those of our simulations in which the ionizing source is one of the 30 000 K or 40 000 K model stars (with the exception of the supergiant model with $0.1\,Z_\odot$ where He iii is the most abundant ionization stage for $\approx 0.05\,r_{\mathrm{H\,ii}}$). Only the dwarf and supergiant stars with effective temperatures of 50 000 K have appreciable He iii Strömgren spheres that reach $\approx 0.05\,r_{\mathrm{H\,ii}}$ (for $2.0\,Z_\odot$) to $\approx 0.10\,r_{\mathrm{H\,ii}}$ (for $0.1\,Z_\odot$).

Although the number densities of metal ions are much smaller than the number densities of hydrogen and helium ions, metals have a large impact on the energy balance of the ionized gas, and interpretation of metal line ratios as markers for the galactic evolution (cf. Balser et al. 2011 and the references therein) requires knowledge of the relation between the ionization fractions of metals and the SEDs of the ionizing sources. For example, the O iii/O ii ratio results from the O ii ionizing flux, which in turn depends not only on the effective temperature of the ionizing sources, but also on their metallicity and their atmospheric density stratification. In the H ii regions around the 30 000 K dwarf stars, O iii is the most abundant ionization stage of oxygen just within less than the innermost $\approx 0.05\,r_{\mathrm{H\,ii}}$ for all metallicities. The O ii-ionizing fluxes of the supergiants at 30 000 K exceed the fluxes of the dwarf stars of the same effective temperature by approximately 3 dex, which results in a more extended volume in which O iii is the dominant ionization stage of oxygen. The extension of this volume strongly depends on metallicity. Its radius $r_{\mathrm{O\,iii}}$ is $\approx 0.4\,r_{\mathrm{H\,ii}}$ for $0.1\,Z_\odot$, but drops to $\approx 0.1\,r_{\mathrm{H\,ii}}$ for a metallicity of $2.0\,Z_\odot$.

The O ii-ionizing fluxes of 40 000 K supergiants, which differ by $\approx 1.3$ dex between the model with $0.1\,Z_\odot$ and $2.0\,Z_\odot$, show a stronger metallicity dependence than the O ii-ionizing fluxes of the dwarf stars with the same effective temperature, which differ by $\approx 0.6$ dex. Consequently, the radii of the O iii dominated parts of the H ii regions around the supergiants vary stronger ($r_{\mathrm{O\,iii}} \approx 0.98\,r_{\mathrm{H\,ii}}$ for $0.1\,Z_\odot$, $r_{\mathrm{O\,iii}} \approx 0.55\,r_{\mathrm{H\,ii}}$ for $2.0\,Z_\odot$) than the radii around the dwarf stars ($r_{\mathrm{O\,iii}} \approx 0.97\,r_{\mathrm{H\,ii}}$ for $0.1\,Z_\odot$, $r_{\mathrm{O\,iii}} \approx 0.87\,r_{\mathrm{H\,ii}}$ for $2.0\,Z_\odot$). O iii is the most abundant ionization stage within the entire Strömgren spheres for dwarf and supergiant stars of all metallicities for an effective temperature of 50 000 K.

Like the O iii fraction relative to the total amount of oxygen, the fraction of S iv decreases for higher metallicities as can be expected in view of the almost identical ionization energies of S iii (2.56 Ryd) and O ii (2.58 Ryd). Still, the relative fraction of S iv is considerably smaller than the relative fraction of O iii within the same H ii region. The reason is the smaller ionization cross-section from the ground-state of S iii ($0.36 \cdot 10^{-18}\,\mathrm{cm}^2$ at the ionization edge) compared to that of O ii ($10.4 \cdot 10^{-18}\,\mathrm{cm}^2$ at the ionization edge).

The above result that low stellar metallicities lead to harder ionizing spectra and thus to a larger fraction of high ionization stages is in agreement with spectroscopic observations. For example, Rubin et al. (2007, 2008) found rising $\langle \mathrm{Ne}^{2+}\rangle/\langle \mathrm{Ne}^{+}\rangle$ and $\langle \mathrm{S}^{3+}\rangle/\langle \mathrm{S}^{2+}\rangle$ ratios for increasing distance from the galactic centers of M 83 and in M 33, based on mid-IR observations with the Spitzer Space telescope. This relation is likely to be connected with the lower metallicities in the outer parts of the galaxies (cf. Rubin et al. 2007). Further observations of the metal-poor galaxy NGC 6822 (Rubin et al. 2012) have found larger fractions of the higher ionization stages than in M 83 (supersolar metallicity) or in M 33 (roughly solar metallicity). These results might, however, additionally be influenced by other factors, like different stellar mass functions or different effective temperatures of the ionizing stars as a function of the chemical composition of the star-forming gas.

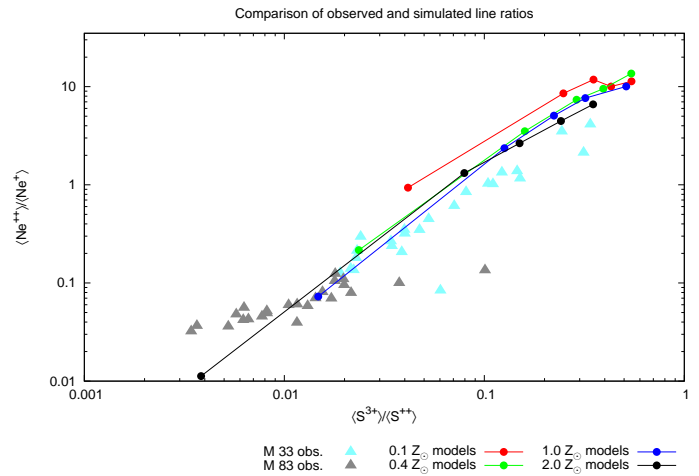

**Fig. 1.** Comparison of observed ionic number ratios $\langle \mathrm{S}^{3+}\rangle/\langle \mathrm{S}^{2+}\rangle$ and $\langle \mathrm{Ne}^{2+}\rangle/\langle \mathrm{Ne}^{+}\rangle$ to the corresponding results from our model H ii regions. The triangles represent H ii regions in the metal rich galaxy M 83 (gray, values from Rubin et al. 2007) and M 33 (cyan triangles, values from Rubin et al. 2008), the circles represent model H ii regions using as ionizing sources dwarf stars with temperatures from 35 000 K to 55 000 K and metallicities of $0.1\,Z_\odot$ to $2.0\,Z_\odot$. The lower metallicities are correlated with a larger fraction of the higher ionization stages both in the observations and in the synthetic H ii region models.

In Figure 1 we compare the simulated $\langle \mathrm{Ne}^{2+}\rangle/\langle \mathrm{Ne}^{+}\rangle$ and $\langle \mathrm{S}^{3+}\rangle/\langle \mathrm{S}^{2+}\rangle$ ratios of our model H ii regions with the corresponding ion ratios determined from the observations described by Rubin et al. (2007, 2008).[14] The figure also shows that the metallicity-dependence of the ionization structure decreases for larger effective temperatures of the ionizing stars as can be expected from the ionizing fluxes shown in Table 4.

**Emission line spectra of H ii regions at different metallicity.** Comparing observed emission line spectra to synthetic nebular models is the most important approach for obtaining information about the temperature, the density, and the ionization structure of H ii regions. To investigate the variations in the emission line ratios, we have computed the fluxes of the collisionally excited optical lines [N ii] 6584 Å + 6548 Å, [O ii] 3726 Å + 3729 Å, [O iii] 5007 Å + 4959 Å, and [S ii] 6716 Å + 6731 Å for each of our H ii region models. In Table 6 the results are shown relative to the corresponding Hβ emission.

In most cases the maximal line emission in the optical range is reached for a metallicity of $0.4\,Z_\odot$. This non-monotonic behavior is the result of two opposing metallicity-dependent effects. On the one hand, a higher metallicity increases the number density of ions that are potential line emitters. On the other hand, the lower temperature of the gas decreases the probability of a collisional excitation process which finally leads to the emission of optical line radiation. The effect of the lower temperatures dominates for $2.0\,Z_\odot$ where in most cases the fluxes for this metallicity are the weakest among the considered metallicities.[15] This is ex-

---

[14] For this comparison we scaled the ionizing fluxes to $10^{49}$ hydrogen-ionizing photons per second and used a hydrogen number density in the gas of $1000\,\mathrm{cm}^{-3}$ to match the assumptions in Rubin et al. (2008).

[15] The cooling by the infrared transitions between the fine-structure levels partly leads to simulated temperatures of the gas which are below 3000 K for a metallicity of $2.0\,Z_\odot$. This result is also obtained with other nebular model codes, e.g., Cloudy (Ferland et al. 2013). Such low temperatures have, however, not been observed in H ii regions. The reason

plained by the equation for the rate coefficients for the collisional excitation,

$$C_{lu} = n_e \left(\frac{2\pi}{kT}\right)^{1/2} \frac{\hbar^2}{m_e^{3/2}} \frac{\Omega_{lu}(T)}{g_l} e^{-h\nu_{lu}/kT} \qquad (5)$$

(Mihalas 1978; $n_e$ is the number density of electrons, $m_e$ the electron mass, $\Omega_{lu}(T)$ the velocity-averaged collision strength, a slowly-varying function of temperature, $h\nu_{lu}$ the transition energy, and $g_l$ the statistical weight of the lower level), which leads to qualitatively different temperature dependences for collisionally excited lines in different wavelength ranges. The exponential term in Eq. 5 rises strongly with increasing temperature if the energy difference between the levels is large compared to $kT$, as is the case for lines in the visible and ultraviolet range. For the lines in the mid-infrared/far-infrared range, the energy difference between the upper and the lower level is significantly smaller than $kT$. Thus, the exponential term is close to unity in the entire temperature range found in H II regions and the rate coefficients are roughly proportional to the inverse square root of the temperature. This implies that the cooling by infrared lines becomes more effective for decreasing temperatures, i.e., there is a positive feedback between the lower temperatures and the infrared line cooling. As a result, a small increase of the metallicity has a strong influence on the temperature structure if it causes the infrared transitions mentioned above to become the dominant cooling processes.

**The influence of metallicity-dependent SEDs on the temperature and ionization structure of gas with fixed metallicity.** In addition to the models where stellar and gas metallicity are equal, we have computed models where the metallicity of the gas is fixed at $1.0 Z_\odot$, but the metallicity of the stellar sources is varied, in order to analyze the effects of stellar SEDs independently of the effects of the chemical composition of the gas. The simulations have been performed using the four different 40 000 K dwarf star models.

The sizes of the hydrogen Strömgren spheres show variations of approximately 5%, which are mainly caused by lower recombination rates in the gas around the lower-metallicity stars where the temperatures are higher than in the gas surrounding the model stars with higher metallicities.[16] The differences are more pronounced for the ionization structure of oxygen and sulfur, especially for the ratios O II/O III and S II/S IV. For the model with a metallicity of $2.0 Z_\odot$, O III is the most abundant ion of oxygen within the inner $\approx 0.85 r_{\text{H\,II}}$, while this is the case for $\approx 0.97 r_{\text{H\,II}}$ at a metallicity of $0.1 Z_\odot$. For S IV the corresponding values are $0.28 r_{\text{H\,II}}$ and $0.59 r_{\text{H\,II}}$, respectively.

The emission of the [O III] 5007 Å + 4959 Å lines increases for decreasing stellar metallicities as a result of the higher temperatures and the larger fraction of O III ions. At a metallicity

---

for this discrepancy is not entirely clear. Shields & Kennicutt (1995) have analyzed the impact of dust within high-metallicity H II regions and attribute the lack of H II regions with very low temperatures to the depletion of metals from the gas phase by the formation of dust and the heating by the emission of electrons from the UV-irradiated dust grains. The spectral properties of H II regions in the metal-rich ($Z \approx 2 Z_\odot$) starforming galaxy M 83 with relatively weak collisionally excited emission features have been described in detail by Bresolin & Kennicutt (2002), who stress that for these conditions a determination of the electron temperature by means of emission line analysis could not be carried out accurately.

[16] The emission of photons whose energies considerably exceed the Lyman-Edge is higher for lower metallicities. Consequently the mean gain of thermal energy per ionization process is larger.

---

of $0.1 Z_\odot$ the emission of these lines is twice as strong as at $2.0 Z_\odot$ (cf. the lower part of Table 6). By contrast, the emission of lines associated with singly ionized atoms, like [N II] 6584 Å + 6548 Å and [O II] 3726 Å + 3729 Å decreases for lower stellar metallicities since then the number of singly ionized ions of these elements is reduced as the shells in which they are the dominant ionization stage become thinner.

### 3.2. 3-dimensional, time-dependent simulations of the ionization structures of the metals in H II regions

We will now abandon the assumption of a homogeneous gas and a spherically symmetric ionized bubble and discuss the effects of inhomogeneities of the gas density. This is motivated by the fact that H II regions are obviously not homogeneous spheres, but are characterized by a more complex structure. The inhomogeneous density structure of H II regions has been treated, for example, by Wood et al. (2004, 2005), who focused on the ability of photons to escape from a "porous" H II region into the diffuse component of the ionized gas in a galaxy, Wood et al. (2013) who studied the consequences of inhomogeneous density structures for the determination of metallicities in H II nebulae, and by Dale & Bonnell (2011), Walch et al. (2012), Dale et al. (2013) who studied the interaction between stellar radiation fields and the density structure of ionized gas. Apart from the metallicity it is therefore the complex geometric structure and the embedded clumps which are of importance for the nebular models and a corresponding comparison of the calculated emission line strengths with the observed spectral features.

To include such effects in our computations, we have developed a 3d radiative transfer code based on the approach by Weber et al. (2013), but considerably extended to account for the time-dependent ionization structure of the metals within the gas. In the description of our method we will first introduce our ray-tracing approach to 3-dimensional radiative transfer. Subsequently we will present a solution of the time-dependent rate equations and treat the influence of the metal ions on the evolution of the temperature structure. Finally, we will describe the generation of synthetic narrowband images, which can link the theoretical models with observations.

This numerical approach will be applied to inhomogeneous, fractally structured gas with various metallicities and sources of ionization. The results of these computations results will be compared with the results obtained for a homogeneous gas distribution. Additionally we will account for the temporal evolution of H II regions and study the effects of the distribution of the sources within clusters of hot stars on the emergent emission line flux of the surrounding ionized gas.

#### 3.2.1. 3-dimensional radiative transfer based on ray tracing

The propagation of light along straight lines directly leads to the concept of a *ray-by-ray solution*, where the luminosity of each source is distributed among a set of rays originating from the source. The main aspects of the ray-by-ray solution are the isotropic distribution of rays around each source and the solution of the radiative transfer equation along each of these rays.

The sources are characterized by a specified spectral energy distribution $F_\nu$ and luminosity $L_\nu(l) = 4\pi R_l^2 F_\nu(l)$ where $R_l$ is the radius of the $l$-th source. The frequency-dependent luminosity $L_\nu$ of each source is distributed evenly[17] among $N_{\text{rays}}$ rays, such that each ray is associated with a luminosity of $\tilde{L}_\nu = L_\nu / N_{\text{rays}}$. The rays are then traced from the source(s) until the border of the simulated volume is reached.

**Table 6.** Comparison of the nebular emission line ratios of H II regions for a grid of H II regions where the ionizing source is in each case one of the model stars. We show the ratio of some important nebular line strengths obtained from spherically symmetric nebular models with varying metallicity to the Hβ emission in these model H II regions.

| Model | $Z/Z_\odot$ | [N II] 6584 + 6548 Å | [O II] 3726 + 3729 Å | [O III] 5007 + 4959 Å | [S II] 6716 + 6731 Å |
|---|---|---|---|---|---|
| \multicolumn Same metallicity for H II region and star (dwarfs) | | | | | |
| D-30 | 0.1 | 0.77 | 1.51 | 0.00 | 0.33 |
| | 0.4 | 1.43 | 1.63 | 0.00 | 0.73 |
| | 1.0 | 1.48 | 0.94 | 0.00 | 0.85 |
| | 2.0 | 0.86 | 0.29 | 0.00 | 0.56 |
| D-35 | 0.1 | 0.44 | 1.59 | 1.66 | 0.15 |
| | 0.4 | 1.06 | 2.45 | 1.63 | 0.38 |
| | 1.0 | 1.29 | 1.70 | 0.68 | 0.49 |
| | 2.0 | 1.06 | 0.56 | 0.05 | 0.38 |
| D-40 | 0.1 | 0.09 | 0.45 | 4.62 | 0.09 |
| | 0.4 | 0.29 | 0.99 | 5.88 | 0.21 |
| | 1.0 | 0.35 | 0.73 | 2.99 | 0.24 |
| | 2.0 | 0.23 | 0.28 | 0.58 | 0.16 |
| D-45 | 0.1 | 0.06 | 0.28 | 5.47 | 0.07 |
| | 0.4 | 0.16 | 0.58 | 8.34 | 0.18 |
| | 1.0 | 0.20 | 0.48 | 4.52 | 0.21 |
| | 2.0 | 0.12 | 0.20 | 1.00 | 0.13 |
| D-50 | 0.1 | 0.04 | 0.21 | 5.98 | 0.07 |
| | 0.4 | 0.12 | 0.43 | 10.07 | 0.18 |
| | 1.0 | 0.15 | 0.41 | 6.57 | 0.21 |
| | 2.0 | 0.09 | 0.22 | 1.88 | 0.13 |
| D-55 | 0.1 | 0.03 | 0.14 | 6.60 | 0.06 |
| | 0.4 | 0.08 | 0.31 | 12.40 | 0.17 |
| | 1.0 | 0.11 | 0.34 | 10.56 | 0.22 |
| | 2.0 | 0.08 | 0.24 | 3.79 | 0.16 |
| \multicolumn Same metallicity for H II region and star (supergiants) | | | | | |
| S-30 | 0.1 | 0.60 | 1.95 | 0.69 | 0.14 |
| | 0.4 | 1.33 | 2.33 | 0.46 | 0.37 |
| | 1.0 | 1.30 | 1.25 | 0.09 | 0.41 |
| | 2.0 | 0.64 | 0.25 | 0.01 | 0.25 |
| S-35 | 0.1 | 0.15 | 0.78 | 3.62 | 0.07 |
| | 0.4 | 0.56 | 1.92 | 3.18 | 0.19 |
| | 1.0 | 1.15 | 1.91 | 0.84 | 0.28 |
| | 2.0 | 1.17 | 0.74 | 0.04 | 0.26 |
| S-40 | 0.1 | 0.07 | 0.35 | 4.99 | 0.06 |
| | 0.4 | 0.20 | 0.79 | 6.25 | 0.14 |
| | 1.0 | 0.48 | 1.16 | 2.13 | 0.18 |
| | 2.0 | 0.97 | 0.87 | 0.15 | 0.21 |
| S-45 | 0.1 | 0.04 | 0.23 | 5.77 | 0.05 |
| | 0.4 | 0.12 | 0.49 | 8.41 | 0.13 |
| | 1.0 | 0.17 | 0.46 | 3.91 | 0.15 |
| | 2.0 | 0.27 | 0.24 | 0.66 | 0.09 |
| S-50 | 0.1 | 0.03 | 0.16 | 6.34 | 0.05 |
| | 0.4 | 0.08 | 0.33 | 10.84 | 0.14 |
| | 1.0 | 0.10 | 0.38 | 7.35 | 0.17 |
| | 2.0 | 0.07 | 0.19 | 2.40 | 0.13 |
| \multicolumn Different metallicity for H II region ($Z = Z_\odot$) and star | | | | | |
| D-40 | 0.1 | 0.24 | 0.49 | 5.11 | 0.30 |
| | 0.4 | 0.29 | 0.57 | 3.42 | 0.25 |
| | 1.0 | 0.35 | 0.73 | 2.99 | 0.24 |
| | 2.0 | 0.43 | 0.93 | 2.42 | 0.25 |

The integrated form of the radiative transfer equation along a ray is

$$I_\nu(s) = I_\nu(s_0(n)) \, e^{\tau_\nu(s_0(n)) - \tau_\nu(s)} + \int_{s_0(n)}^{s} \eta_\nu(s') \, e^{\tau_\nu(s') - \tau_\nu(s)} \, \mathrm{d}s', \quad (6)$$

where $I_\nu(s)$ is the intensity at the position $s$, $s_0(n)$ corresponds to the starting point of each ray $n$, $\eta_\nu(s')$ is the emissivity of the gas[18] at the position $s'$, and $\tau_\nu(s) = \int_{s_0(n)}^{s} \chi_\nu(s') \, \mathrm{d}s'$ is the optical depth with respect to the source.

The simulated volume is discretized into a Cartesian grid of cells and each of the sources is located in the center of one of these cells. The energy deposited per time in the cells crossed by a ray $n$ along the distance $s_n(m)$ between the source and the entry point into a cell $m$ is then calculated as

$$\Delta\dot{E}(s_n(m)) = \int_0^\infty \tilde{L}_\nu(s_0(n)) \left(1 - e^{-\tau_\nu(s_n(m))}\right) \mathrm{d}\nu \quad (7)$$

(for details see Weber et al. 2013).

**Temporal evolution of the ionization structure of metals.**
Modeling the evolution of the ionization structures is not straightforward. The rate equations form a stiff set of differential equations as the timescales for the various ionization and recombination processes in the cell may differ by several orders of magnitude. Furthermore the timescales differ between various cells depending on whether they are passed by an ionization front and how distant they are from the sources of ionization. However, the eigenvalue approach presented here provides a novel method to solve the rate equations in a stable and efficient way.

The solution

$$\boldsymbol{x}(t) = \boldsymbol{x}^\infty + \boldsymbol{x}_{\text{hom}}(t) \quad (8)$$

of the inhomogeneous system of differential equations defined by Eq. 3 is composed of the "equilibrium solution" $\boldsymbol{x}^\infty$ of Eq. 4 and the solution of the corresponding homogeneous system

$$\boldsymbol{E}' \cdot \frac{\mathrm{d}}{\mathrm{d}t} \boldsymbol{x}_{\text{hom}}(t) + \boldsymbol{G}' \cdot \boldsymbol{x}_{\text{hom}}(t) = \boldsymbol{0}, \quad (9)$$

which is written in components as

$$\begin{pmatrix} \frac{\mathrm{d}x_{\text{hom},1}}{\mathrm{d}t} \\ \vdots \\ \frac{\mathrm{d}x_{\text{hom},N-1}}{\mathrm{d}t} \\ 0 \end{pmatrix} + \begin{pmatrix} g'_{1,1} & \cdots & g'_{1,N-1} & 0 \\ \vdots & \ddots & \ddots & \vdots \\ g'_{N-1,1} & \cdots & g'_{N-1,N-1} & 0 \\ 1 & \cdots & 1 & 1 \end{pmatrix} \cdot \begin{pmatrix} x_{\text{hom},1} \\ \vdots \\ x_{\text{hom},N-1} \\ x_{\text{hom},N} \end{pmatrix} (t) = \begin{pmatrix} 0 \\ 0 \\ \vdots \\ 0 \end{pmatrix}. \quad (10)$$

---

[17] To ensure that each of the rays represents a solid angle of the same size we use the "HEALPiX" method described by Górski et al. (2005) for the distribution of the rays.

[18] In the 3-dimensional simulations we apply the on-the-spot approximation (Zanstra 1931, Baker & Menzel 1938, Spitzer 1998) for hydrogen and helium to account for the diffuse radiation field. This approximation assumes that photons generated by a recombination process which directly leads into the ground state of the recombined particle are locally re-absorbed. Thus, the diffuse ionization field is not considered explicitly; instead, the rates for recombination processes that directly lead into the ground state are subtracted from the total recombination rates.

In Eq. 10 the condition of particle conservation is as an example inserted for $k = N$. The particle conservation implies that the sum of all components of a solution of the homogeneous system of differential equations is 0. For the solution $x_{hom}(t)$ of Eq. 10, the $k$-th component is thus computed as

$$x_{hom,k}(t) = - \sum_{i=1, i \neq k}^{N} x_{hom,i}(t). \tag{11}$$

The general structure of the solution for the components of the homogeneous equation different from $k$ is

$$x_{hom,i}(t) = \sum_{j=1}^{N-1} v_{i,j} e^{-\delta_j(t-t_0)} \gamma_j \tag{12}$$

where $v_{i,j}$ is the $i$-th component of the $j$-th eigenvector of the system of differential equations obtained from Eq. 10 by removing the $k$-th line of $x$ and the $k$-th line and $k$-th column of $G'$. $\delta_j$ are the corresponding eigenvalues. The values of $\gamma_j$ are defined by the occupation numbers at $t_0$, the beginning of the respective timestep.

Equations 11 and 12 are combined in matrix-vector notation using a matrix $V$ which is composed of the eigenvectors $v_j$, but additionally contains in its $k$-th row and $j$-th column the negative of the scalar product $\mathbf{1} \cdot v_j$ (where the vector $\mathbf{1}$ contains the entry 1 in each of its components):

$$x_{hom}(t) = V \cdot \delta(t) \cdot \gamma = \begin{pmatrix} v_1 & v_2 & \dots & v_{N-1} \\ -\mathbf{1} \cdot v_1 & -\mathbf{1} \cdot v_2 & \dots & -\mathbf{1} \cdot v_{N-1} \end{pmatrix} \cdot \delta(t) \cdot \gamma. \tag{13}$$

$\delta(t)$ is a diagonal matrix with the entries $\delta_{j,j} = e^{-\delta_j(t-t_0)}$ and $\gamma$ is a vector which contains in its $j$-th row the coefficient $\gamma_j$ (in Eq. 13 we set $k = N$ again).

The vector $\gamma$ is determined by inserting the occupation numbers for $t = t_0$ which are known from the previous timestep. In this case $\delta$ becomes the identity matrix and from Eq. 12 follows

$$\tilde{V} \cdot \gamma = \tilde{x}_{hom}(t_0) = \tilde{x}(t_0) - \tilde{x}^{\infty} \tag{14}$$

where $\tilde{V}$ corresponds to the matrix $V$ without the $k$-th row (the $k$-th row is also removed for $\tilde{x}_{hom}$ and $\tilde{x}^{\infty}$). Eq. 14 is solved for $\gamma$ by

$$\gamma = \tilde{V}^{-1} (\tilde{x}_{hom}(t_0) - \tilde{x}^{\infty}.) \tag{15}$$

With Eqs. 15 and 13 , the values for $x_{hom}(t)$

$$x_{hom}(t) = V \cdot \delta(t) \cdot \tilde{V}^{-1} (\tilde{x}(t_0) - \tilde{x}^{\infty}) \tag{16}$$

are obtained. The result for $x(t)$ is the initial value for the next timestep and after the re-computation of the rate coefficients the matrix $G'$ is updated. We currently consider the ionization stages I and II of hydrogen, I to III of helium, and I to IV of carbon, nitrogen, oxygen, neon, and sulfur, but in principle the method can be extended to any number of ionization stages and any set of elements required to describe the respective problem. The presented method has proven to be numerically very stable. Particle conservation is preserved for each element with a relative deviation of less than $10^{-8}$.

In the next paragraphs we will describe the computation of the rate coefficients for the recombination and ionization processes which via the rate matrix $G'$ determine the temporal evolution of the ionization fractions in $x(t)$.

**Recombination Rates.** The relevant recombination processes in the interstellar gas are radiative and dielectronic recombination. In radiative recombination processes a free electron is captured by an ion and its kinetic and potential energy (relative to the bound state immediately after the recombination) is converted into the energy of the emitted photon. In dielectronic processes the energy of the captured electrons excites another electron of the ion, resulting in a doubly excited intermediate state. Dielectronic recombinations can only occur for discrete electron energies because of the additional bound-bound-process. The recombination rate coefficient of an ion $j$ in a grid cell $m$ filled with gas at a temperature of $T(m)$ is given by $\mathcal{R}_{j,j-1}(m) = \alpha_j(T(m)) \cdot n_e(m)$, where $n_e(m)$ is the number density of electrons and $\alpha_j(T(m))$ is the total recombination coefficient, composed of a radiative contribution $\alpha_{r,j}(T(m))$ and a dielectronic contribution $\alpha_{d,j}(T(m))$:

$$\alpha_j(T(m)) = \alpha_{r,j}(T(m)) + \alpha_{d,j}(T(m)). \tag{17}$$

For the temperature-dependent radiative recombination coefficients of metal ions of type $j$ in cell $m$, we use the approximate formula

$$\alpha_{r,j}(T(m)) = A_j T'^{-X_j}(m), \tag{18}$$

where $T'(m) = T(m)/10\,000$ K is the temperature of the gas in units of 10 000 K. The fit parameters $A_j$ and $X_j$ are taken from Aldrovandi & Pequignot (1973), Aldrovandi & Pequignot (1976), Shull & van Steenberg (1982), and Arnaud & Rothenflug (1985).

The dielectronic recombination rates for low temperatures (where $kT$ is smaller than the energy to create the excited intermediate states – as it is typically the case in H ɪɪ regions), are described by the fit formula (Nussbaumer & Storey 1983)[19]

$$\alpha_{dl,j}(T(m)) = 10^{-12} \cdot \left( \frac{a_j}{T'(m)} + b_j + c_j T'(m) + d_j T'^2(m) \right) \cdot T'^{-3/2}(m) e^{-f_j/T'(m)} \text{cm}^3 \text{ s}^{-1}. \tag{19}$$

Here $a_j, b_j, c_j, d_j$, and $f_j$ are fit parameters which depend on the type of the ion.

An additional contribution to the dielectronic recombination rate that becomes relevant for higher temperatures where $kT$ is in the order of the excitation energy has been described by (Burgess 1964) and is approximated by

$$\alpha_{dh,j}(T(m)) = B_j \left( \frac{T}{K} \right)^{-3/2} (m) e^{-T_{0,j}/T(m)} \left( 1 + C_j e^{-T_{1,j}/T(m)} \right). \tag{20}$$

The values for the fit parameters $B_j, C_j, T_{0,j}$, and $T_{1,j}$ were obtained from from the same sources as the radiative recombination coefficients.

The total dielectronic recombination rate $\alpha_{d,j}$ is obtained by adding up the low-temperature and the high-temperature contributions (Storey 1983).

The fit functions for the recombination rates already consider the different recombination paths which finally lead to the ground state of the recombined ion. The recombination rates for hydrogen and helium are computed by interpolating the tables provided by Hummer (1994) and Hummer & Storey (1998).

---

[19] An online collection of the radiative and dielectronic recombination data of metal ions that contains the values used in this work is provided by D. A. Verner, http://www.pa.uky.edu/~verner/rec.html.

**Photoionization rates and computation of the mean intensity.**
The photoionization rates $R_{lu}(m)$ from a level $l$ in one ionization stage to a level $u$ in the next ionization stage are calculated as

$$R_{lu}(m) = \int_{\nu_{lu}}^{\infty} \frac{4\pi a_{lu}(\nu)}{h\nu} J_\nu(m) \, d\nu, \tag{21}$$

where $J_\nu(m)$ is the mean intensity, $\nu_{lu}$ the threshold frequency for the considered ionization process and $a_{lu}(\nu)$ the frequency-dependent ionization cross-section.[20]

To determine the average intensity $J_\nu$ needed to compute the radiative rates, we proceed as follows. From our discussion of the radiative transfer we already know the total energy absorbed (per unit time) by a cell, namely that given by Eq. 7. The number of photons absorbed per unit time in the cell in a particular transition is given by the same integral, with the integrand divided by the energy $h\nu$ of a photon and weighted by the relative contribution of that transition to the total opacity,

$$\sum_{\substack{\text{ray} \\ \text{segments}}} \int \frac{\tilde{L}_\nu^{(\text{inc.})}(1 - e^{-\tau_\nu})}{h\nu} \frac{\chi_{lu}(\nu)}{\chi_\nu^{\text{tot}}} \, d\nu = \tag{24}$$
$$= V n_l R_{lu} = V n_l \int \frac{4\pi a_{lu}(\nu)}{h\nu} J_\nu \, d\nu,$$

where $\tilde{L}_\nu^{(\text{inc.})}$ is again (for every ray passing through that cell) the luminosity of the ray incident on the cell, $\tau_\nu$ is the total optical depth of the cell along that ray, and $V$ is the volume of the cell. Since $\chi_{lu}(\nu)$ is simply $n_l a_{lu}(\nu)$, we see that the expression for $J_\nu$ that ensures consistency between radiative transfer and rate equations is

$$J_\nu = \frac{1}{4\pi V} \sum_{\substack{\text{ray} \\ \text{segments}}} \frac{\tilde{L}_\nu^{(\text{inc.})}(1 - e^{-\tau_\nu})}{\chi_\nu^{\text{tot}}} \tag{25}$$

which (as it must be) is of course independent of the actual transition considered in the discussion above, and can be used to compute the photoionization rates of all elements and ionization stages (for more details see Weber et al. 2013).

---

[20] For the computation of the frequency-dependent ionization cross-section, we use the "Seaton approximation" (Seaton 1958)

$$a_{lu}(\nu) = a_{0,lu} \left( \beta_{lu} \left( \frac{\nu}{\nu_{lu}} \right)^{-s_{lu}} + (1 - \beta_{lu}) \left( \frac{\nu}{\nu_{lu}} \right)^{-s_{lu}-1} \right), \tag{22}$$

where $a_{0,lu}$, $\beta_{lu}$, and $s_{lu}$ are fit parameters for the numerical results of the quantum-mechanical calculations of the cross-sections as a function of photon energy. The nebular approximation is applied and only ionization processes from atoms in the ground state are considered, but we note that there may be several ionization channels with different electronic states of the ionization product. The total ionization rate in this case is $\mathcal{R}_{i,j} \approx \sum_{u=l_0}^{u_{max}} R_{l_0,u}$, where $l_0$ is the ground-level of ion $i$ and the summation is carried out over the considered upper levels.
In addition to the radiative ionization rates we also compute the collisional ionization rate coefficients by applying an approximation from Seaton as given by Mihalas (1978),

$$C_{lu} = n_e \frac{1.55 \cdot 10^{13}}{(T/K)^{1/2}} \bar{g} \frac{e^{-h\nu_{lu}/kT}}{h\nu_{lu}/kT}, \tag{23}$$

where $\bar{g}$ takes the value 0.1, 0.2, or 0.3 for an initial ionic charge of 0, 1, or $\geq 2$, respectively.

**The temperature structure of evolving H II regions.** The assumption that in an H II region the heating and the cooling rates are in balance (as it has been done in the time-independent spherically symmetric approach presented in Sect. 3.1) and the temperature of the gas consequently remains constant is only valid for a steady-state H II region. However, in evolving H II regions the heating and cooling rates are in general not equal and the variation of the thermal energy content within the considered volume elements of the simulated gas has to be taken into account explicitly. Our simulations account for the photoionization of hydrogen and helium as heating processes. The photoionization heating rate (per volume) from a state $l$ into the state $u$ is

$$\Gamma_{lu}(m) = n_l \int_{\nu_{lu}}^{\infty} \frac{4\pi a_{lu}(\nu) h(\nu - \nu_{lu})}{h\nu} J_\nu(m) \, d\nu, \tag{26}$$

where $n_l$ is the occupation number of the lower state $l$ and $h\nu_0$ the ionization energy. Again we assume the nebular approximation that all ionization processes take place from the ground state of an ion.

The considered cooling processes are the radiative recombination of hydrogen (Hummer 1994) and helium (Hummer 1994, Hummer & Storey 1998) as well as the radiative decay of collisionally excited ions.[21] Furthermore, the simulations account for the cooling by free-free radiation (Osterbrock & Ferland 2006).

The heating and cooling rates are used to compute the change of the thermal energy content of the gas within a grid cell $m$. The temperature $T(m)$ of the gas in a cell $m$ is computed from the total content of thermal energy $E_{\text{therm}}(m)$ in that cell by

$$\frac{3}{2} N_{\text{part}}(m) kT(m) = \frac{3}{2} V n_{\text{part}}(m) kT(m) = E_{\text{therm}}(m),$$

$$T(m) = \frac{2 E_{\text{therm}}(m)}{3 V n_{\text{part}}(m) k}. \tag{27}$$

Here $N_{\text{part}}(m)$ is the number of gas particles (electrons, atoms and ions) within a grid cell, $n_{\text{part}}(m)$ the number density of the particles and $V(m)$ the volume of the grid cell. After each timestep $\Delta t$, the thermal energy $E_{\text{therm}}^{\text{new}}(m)$ is re-computed as

$$E_{\text{therm}}^{\text{new}}(m) = E_{\text{therm}}^{\text{old}}(m) + (\Gamma(J_\nu(m)) - \Lambda(n_e(m), T(m))) \cdot V \cdot \Delta t, \tag{28}$$

where

$$\Gamma(J_\nu(m)) = \sum_i n_i \Gamma_i(J_\nu(m))$$

and

$$\Lambda(n_e(m), T(m)) = \Lambda_{\text{ff}}(n_e(m), T(m)) + \sum_i n_i \Lambda_i(n_e(m), T(m))$$

are the total heating and cooling rates per volume unit. The summations are carried out over all ionization stages $i$, where $n_i$ are the number densities of the ions. $\Gamma_i(J_\nu(m))$ and $\Lambda_i(n_e(m), T(m))$ are the respective heating and cooling rate coefficients, which in turn depend on the radiation field described by the mean intensity $J_\nu$, on the electron density $n_e$ and the temperature $T$ within a given grid cell. $\Lambda_{\text{ff}}$ is the free-free cooling rate, $\Delta t$ is the length of a timestep and $E_{\text{therm}}^{\text{old}}(m)$ the energy content of the cell $m$ before the timestep. $\Gamma_i$ and $\Lambda_i$ already include the contributions of the different heating and cooling processes connected to an ionization stage $i$.

---

[21] For references concerning the collisional processes in the relevant ions, cf. Sect. 3.1.1. (The values for neutral hydrogen were obtained from Anderson et al. 2000.)

Computing synthetic images. Images of gaseous nebulae show 2-dimensional projections of the 3-dimensional emission pattern of these objects. To link the results of our simulations with possible observations we create synthetic images which can be compared with narrow-band images taken in the wavelength range around diagnostically important emission lines.

The dilute gas found in H ɪɪ regions is almost transparent for the radiation of lines in the visible or infrared part of the spectrum if they are either emitted during the transition to a non-ground-level state (e.g., the lines of the Balmer series of hydrogen) or by a forbidden transition into a ground-level state (e.g., the forbidden line of O ɪɪ at 3729 Å). If the absorption and scattering terms are negligible, the intensity of the radiation at a given wavelength is given by

$$I_\nu = \int_{s_c}^{s_d} \eta_\nu(s)\,\mathrm{d}s. \tag{29}$$

Here $\eta_\nu$ is the emissivity of the medium and the integration over $s$ along the line of sight is carried out between the point $s_c$ of the emitting volume which is the closest to the observer and the point $s_d$ which is the most distant from the observer. We consider the frequency-integrated intensities $\bar{I}$ and emissivities[22] $\bar{\eta}$ of the examined emission lines since the wavelength resolution in the 3-dimensional radiative transfer is not sufficient to resolve the line profiles. The number of emitted photons $\mathrm{d}N$ per solid angle $\mathrm{d}\Omega$, detector surface $\mathrm{d}A$, and time $\mathrm{d}t$ therefore is

$$\frac{\mathrm{d}^3 N}{\mathrm{d}\Omega\,\mathrm{d}A\,\mathrm{d}t} \approx \frac{\bar{I}}{h\nu_0} = \int_{s_c}^{s_d} \frac{\bar{\eta}(s)}{h\nu_0}\,\mathrm{d}s, \tag{30}$$

for each of the considered lines, where $\nu_0$ is the frequency of the center of the line. In our discretization scheme the integration along the line of sight is replaced by a summation of the emissivities multiplied with the lengths $\Delta s(m)$ of the ray segments through the cells.

The small-angle approximation that the "rays" connecting the observer and all parts of the emitting regions are parallel can be used if the emission regions are small compared to their distance from the observer (as we assume in our simulations) . In our case the summation is carried out along one of the coordinate axes (cf. Fig. 2).

### 3.2.2. Applications of the 3d approach for the simulation of H ɪɪ regions around hot stars

The presented 3d approach is now applied to examine various aspects of the interaction between hot stars and inhomogeneous H ɪɪ regions which have been sketched in the introduction of Sect. 3 .

Comparison between homogeneous and inhomogeneous H ɪɪ regions. First we use our procedure to simulate the interaction of the radiation field of hot stars and the inhomogeneous H ɪɪ regions surrounding these stars. The inhomogeneous H ɪɪ models are based on a fractal density distribution similar to the distributions that have been used to describe the interstellar gas by Elmegreen & Falgarone (1996) and by Wood et al. (2005). The results are then compared with the ionization structures and emission properties of homogeneous H ɪɪ regions with the same mean densities, metallicities and sources of ionization.

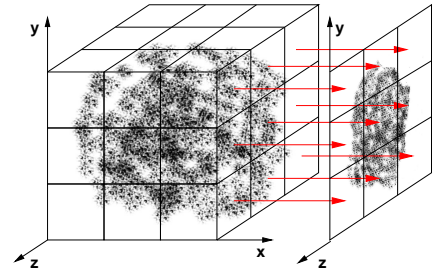

**Fig. 2.** 2-dimensional projections of the calculated 3-dimensional structures represent "synthetic images". The synthetic images of the strengths of emission lines are generated by integrating the emissivities along one of the coordinate axes (in the picture shown along the $x$-axis).

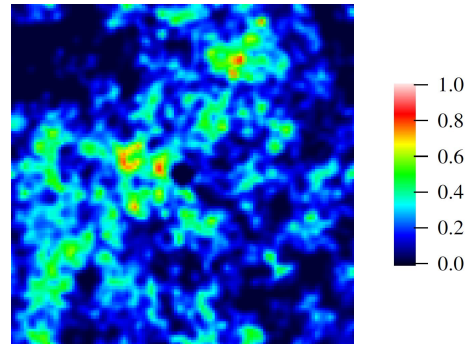

**Fig. 3.** Cross-section through a region of inhomogeneously distributed gas with a fractal density distribution (scaled to 50 hydrogen atoms per cm$^{-3}$) which is used for the 3d simulations shown in Figs. 4 to 9. The volume has a size of (40 pc)$^3$ and is resolved into 101$^3$ grid cells.

In the inhomogeneous models, the mean number density of hydrogen is set to 9 cm$^{-3}$ for the fractal structures in the simulations. Additionally, the simulated gases contains a homogeneous fraction of 1 cm$^{-3}$. The total number densities $n_{\rm H}$ of hydrogen atoms vary between 1 cm$^{-3}$ (where the contribution of the fractal density field is zero) and 93 cm$^{-3}$. The density distribution of the gas is furthermore characterized by the clumping factor $f_{cl} = \langle n_{\rm H}^2 \rangle / \langle n_{\rm H} \rangle^2$ of $\approx 1.75$ and the standard deviation of the hydrogen number density $(\langle n_{\rm H}^2 \rangle - \langle n_{\rm H} \rangle^2)^{1/2} = 8.7$ cm$^{-3}$. In Fig. 3 we show the number density of hydrogen for the mid-plane of the simulated volume, which in the presented simulations contains the source or sources of ionization. As the recombination and collisional excitation rates are proportional to the product of the number densities of the corresponding ions and the number density of electrons, they are approximately proportional to the square of the hydrogen number density in the ionized volumes. In the simulated inhomogeneous volume of gas the line emission per volume unit of the gas thus differs by more than 3 dex for a given temperature and number ratio of the ions. For this density structure we perform simulations of H ɪɪ regions around 40 000 K dwarf stars with metallicities of 0.1 $Z_\odot$ and 1.0 $Z_\odot$ where the relative metallicity in the gas of the H ɪɪ regions is the same as in the stellar atmospheres and the metallicity is assumed to be constant within the simulated volume.

The ionization fractions within the simulated volume, the total emission of some diagnostically important optical lines and the maximal intensities for these lines in the synthetic images of the simulated H II regions are compared in Table 7 . We find that the ratio between the number fractions of ionized hydrogen in the homogeneous and in the inhomogeneous models is approximately 1.6, which roughly agrees with the clumping factor. This behavior is a result of the proportionality of the recombination rates and the square of the gas density. By definition the mean square of the hydrogen density in the inhomogeneous case exceeds the mean square of the hydrogen density in the homogeneous case by the clumping factor. Consequently, a smaller number fraction of ionized hydrogen results for identical sources of ionization. Likewise the clumping increases the neutral fraction for the other considered elements.

---

[22] The frequency-integrated emissivity of a line within a cell is computed as $\bar{\eta}(m) = \frac{1}{4\pi} \frac{\mathrm{d}E(m)}{\mathrm{d}t\,\mathrm{d}V}$. Here $\mathrm{d}E(m)$ is the energy of the line emission in the time interval $\mathrm{d}t$ per volume $\mathrm{d}V$. The numerical values for the emission the recombination lines of hydrogen and helium are based on data given in Osterbrock & Ferland (2006), while the computation of the emission of collisionally excited lines corresponds to the computation of the cooling by these lines as described in Sect. 3.1.1.

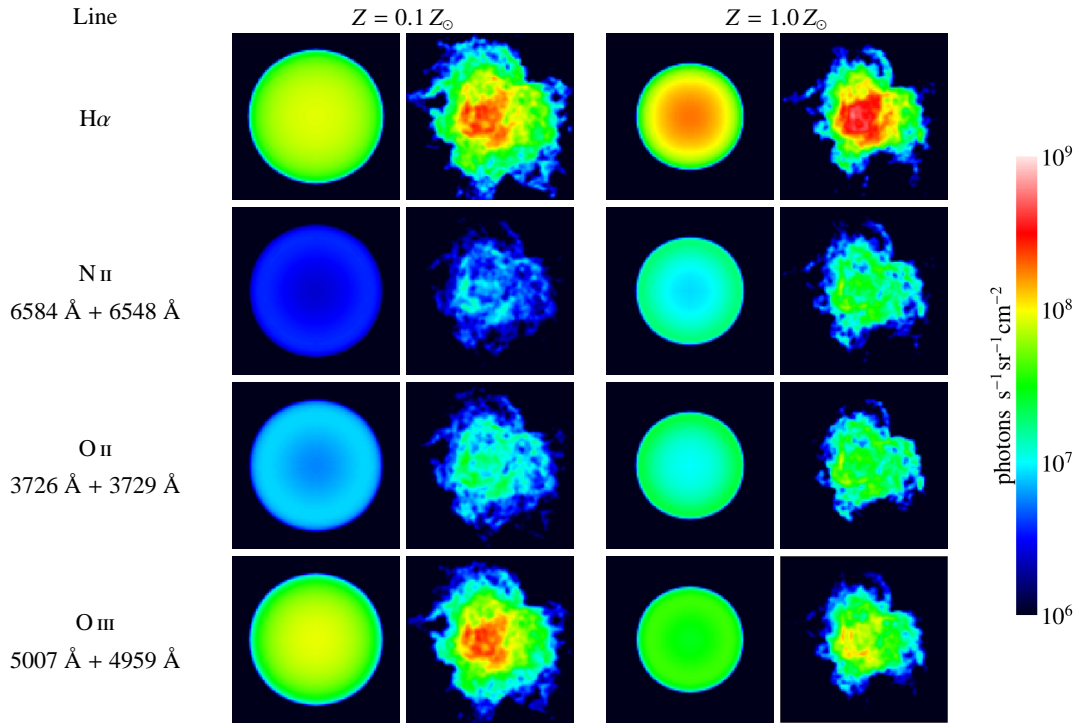

**Fig. 4.** Synthetic images indicating the distribution of the emission line strengths of Hα, [N II] 6584 Å + 6548 Å, [O II] 3726 Å + 3729 Å, and [O III] 5007 Å + 4959 Å for two different values of the metallicity (0.1 $Z_\odot$ and 1.0 $Z_\odot$). Compared are cases where the gas is spherically symmetric structured with a homogeneous density structure ($n_H = 10$ cm$^{-3}$) and cases where the gas is fractally structured with an inhomogeneous density structure of the same mean density. The source of ionization is located in the center of the simulated volume ([40 pc]$^3$) and is represented by the respective 40 000 K dwarf star (D-40) at the same metallicity as the gas (cf. Sect. 2.2).

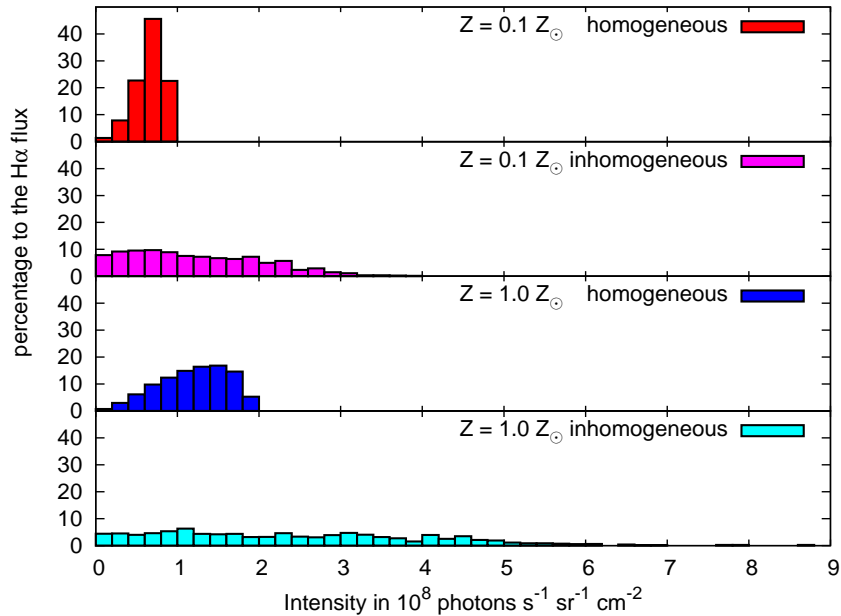

**Fig. 5.** The synthetic images based on the fractally structured H II regions shown in Fig. 4 reveal not only considerably different distributions of the emission line strengths, but also of the intensity patterns compared to those of the corresponding homogeneously structured H II regions (results of the simulations based on 0.1 $Z_\odot$ are presented in red and magenta, those which are based on 1.0 $Z_\odot$ are presented in blue and cyan). The histograms illustrate this behavior for the Hα line by showing the bin-wise integrated photon-numbers of the lines of sight on the abscissa (the intensity range has been subdivided into 50 equidistant bins) and the relative bin-wise percentages to the total Hα flux which is obtained by adding the contributions of those lines of sight which are associated with a specific bin on the ordinate. While for the homogeneous gas (red/blue) the photon distribution is characterized by a pronounced peak near its maximal value – reached at the line of sight crossing the center of the H II region –, a broader distribution is obtained for the inhomogeneous models (magenta/cyan) – here the maximal values are approximately a factor of 4 larger than those obtained for the homogeneous models (the lines of sight which cross the dense clumps of the gas are responsible for this behavior). The histograms reveal further that the maximal photon-numbers of the 0.1 $Z_\odot$ models are lower than those of the corresponding 1.0 $Z_\odot$ models. This behavior is due to the lower temperature levels that are obtained for gases of higher metallicities and which result in higher recombination rates and thus in smaller ionized volumes (cf. Sect. 3.1.2).

Fig. 4 shows the projected images of the resulting photon emission rates $\bar{I}/h\nu_0$ of the H$\alpha$ line and of the collisionally excited lines of [N II] 6584 Å + 6548 Å, [O II] 3726 Å + 3729 Å and [O III] 5007 Å + 4959 Å.

In the homogeneous case the density of the electrons and hydrogen ions varies only marginally over most of the ionized volume. The emission of the H$\alpha$ line, which in the temperature range found in typical H II regions is primarily a recombination line, consequently shows little structure apart from a decrease from the center to the border of the Strömgren sphere, which is a geometrical effect – the length of a path through a sphere decreases for larger impact parameters with respect to the source. By contrast, N II and O II primarily exist within the outer regions of the hydrogen Strömgren sphere. In the projected image of the line emission, this leads to a ring-like distribution of the projected emission line strengths. Fig. 4 shows that for the inhomogeneous density structures the emission is more peaked around those lines of sight that intersect the density maxima ("clumps").

How regions with different intensities in Fig. 4 contribute to the total flux is shown in the histograms of Fig. 5, using the H$\alpha$ line emission as an example. The histogram covers the range for the values of $\bar{I}/h\nu_0$ which appear in our synthetic images with 50 equidistant bins from 0 to $9 \cdot 10^8$ photons s$^{-1}$ sr$^{-1}$ cm$^{-2}$. For each bin we add up the percentages to the total fluxes for all pixels where the intensities are within the corresponding intervals. In the homogeneous models the highest H$\alpha$ intensities are reached for the lines of sight around the center of the ionized volume, while for the inhomogeneous models there is a broader distribution of the intensities. The histogram shows that for the inhomogeneous models the largest intensities exceed the largest intensities for the homogeneous case by a factor of about 4 to 5. Similar factors occur also for the maximal intensities of the considered collisionally excited lines as Table 7 shows. Moreover, the contribution of high intensities for $1.0 Z_\odot$ is larger than for $0.1 Z_\odot$ in the homogeneous and in the inhomogeneous case. The reason is that the recombination coefficient is larger and the ionized volume smaller due to the lower temperatures for the larger metallicity (cf. Sect. 3.1.2), which leads to a reduced ionized volume. The total H$\alpha$ emission is however nearly proportional to the number of ionization processes of hydrogen, which are almost the same in both models. As the same emission occurs in a smaller volume, larger intensities result. (The small differences between the hydrogen ionizing fluxes of the D-40 models at $0.1 Z_\odot$ and $1.0 Z_\odot$ are of minor importance.)

Temporal behavior of the ionization and temperature structure in model H II regions   We next investigate the time-dependence of the ionization and the temperature structure of the gas during the expansion of H II regions into previously neutral interstellar gas with a fractal density structure. These simulations have been performed for the 40 000 K dwarf stars with metallicities of $0.1 Z_\odot$ and $1.0 Z_\odot$ (the metallicities within the gas of the H II regions match the metallicities in the stellar atmospheres). In these simulations the total mass density of the gas in each cell is assumed to remain constant.[23]

The results are presented in Figs. 6 and 7 . In both cases the temperature increases quickly during the ionization process of previously neutral gas crossed by the ionization front. The heating rate is proportional to the number of ionization processes for a given ionizing spectrum. Therefore the heating rate is increased until the gas is ionized to a large extent, such that it becomes optically thin for the ionizing radiation. Unlike the heating rates, the relevant cooling rates are approximately proportional to the square of the electron density, i.e., they continuously increase during the ionization process. Consequently the cooling rates reach their maximum later than the heating rates and the rise of the temperature of the gas during the ionization process is followed by a decrease until the equilibrium temperature is reached. In the inner parts of the ionized region where the ionization timescale and consequently the heating timescale are considerably shorter than the recombination and cooling timescale, the temperature maximum reached depends on the ionizing SED rather than on the metallicity-dependent efficiency of the cooling within the gas. The "overshooting" effect of the temperature is more pronounced for the simulation of the H II region with solar abundance presented in Fig. 7 than for the metal-poor region in Fig. 6 because of the lower equilibrium temperature for the larger metallicity. When the ionized volume is close to its equilibrium value and the outer parts of the equilibrium Strömgren volume are ionized, the ionization process is slower than for the inner regions of the H II region due to the more dilute radiation field and the timescale for the heating process becomes comparable to the cooling timescale. As a result, the value of the "temperature peak" relative to the equilibrium value is lower than in the inner regions of the H II region.

The overshooting of the temperatures affects the emission spectrum of the expanding nebula such that the intensities of collisionally excited lines like the optical [O II] and [O III] lines decrease when the gas cools down. The effect is more pronounced for the gas with $1.0 Z_\odot$ (Fig. 7) than for the gas with $0.1 Z_\odot$ (Fig. 6) as it is the case for the overshooting of the temperature.

A comparison of the temperature and the density of the ionized gas in the first two rows of Fig. 7 shows that the cooling process is faster for denser gas. For given temperature and ionization fractions, the energy content rises linearly with the density of the gas. On the contrary, the cooling rates are approximately proportional to the square of the density and the cooling timescale is therefore roughly proportional to the inverse of the gas density.

Simulations of interstellar gas irradiated by clusters of hot stars.   Finally, we want to see how these results change if we replace the single star in the previous models by a small cluster of stars and study the effects of a different spatial distribution of the stellar sources. The aim of this comparison is to get a rough idea of how the spatial distribution influences the total nebular emission, and we therefore simply use single stars as ionizing

---

[23]   In our simulations we do not consider the advection of matter, i.e., we neglect velocity fields in the H II regions. Furthermore we assume that the sources are created instantaneously, i.e., we do not account for stellar evolution. While stellar evolution processes are obviously not negligible for actual H II regions, the simulations using this approximation nevertheless show the general behavior of gas crossed by an ionization front.

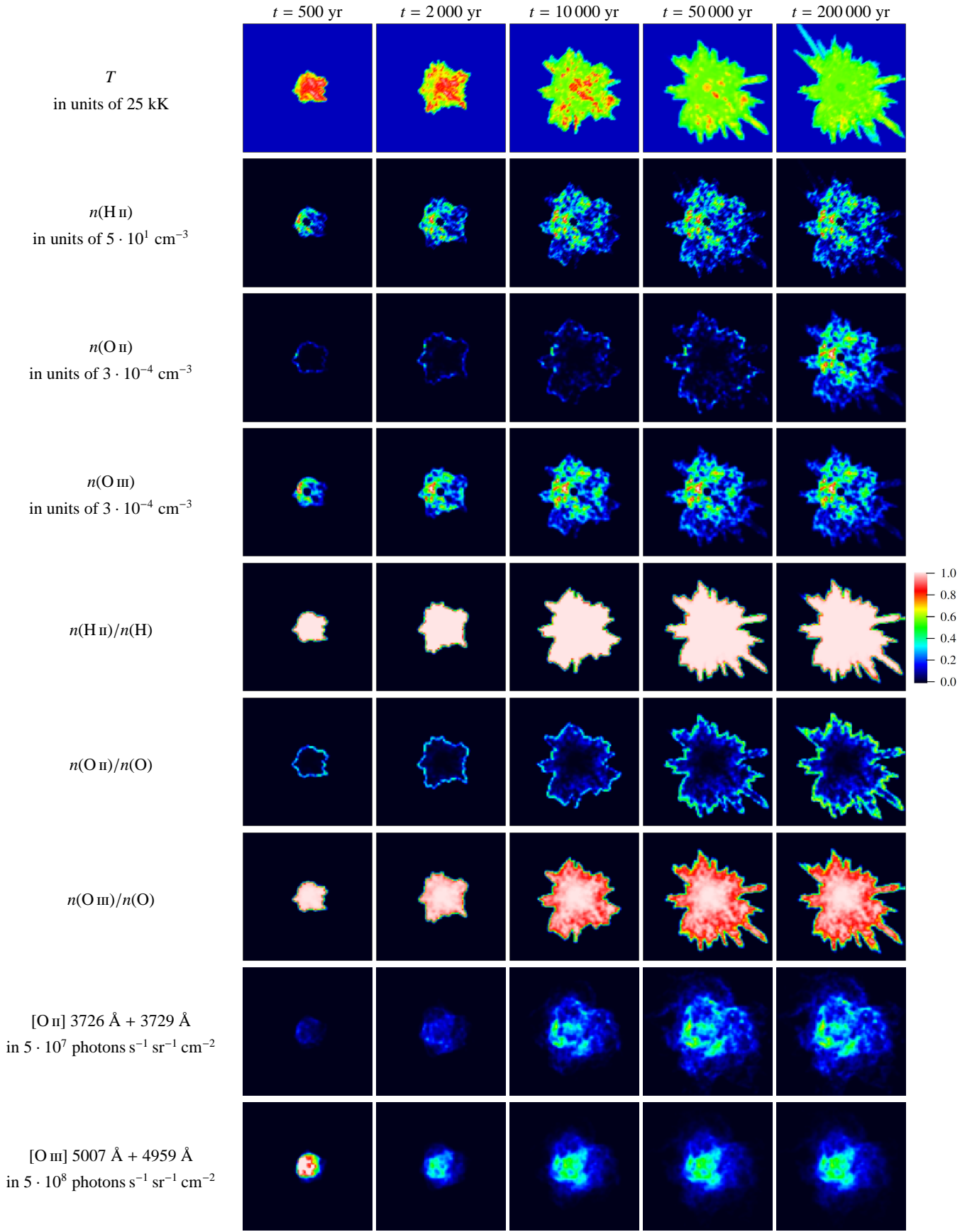

**Fig. 6.** Temporal evolution of an H II region expanding in a volume of $(40 \, \mathrm{pc})^3$ filled with inhomogeneously distributed (initially neutral) gas. The top seven rows show cross-sections through the center of the volume and display the temperature and the absolute as well as relative number densities of the ions H II, O II, and O III. The lower two rows show synthetic images of the line intensities of [O II] 3726 Å + 3729 Å, and [O III] 5007 Å + 4959 Å. In these models, both the central ionizing star (a 40 000 K dwarf) and the gas have a metallicity of $0.1 \, Z_\odot$.

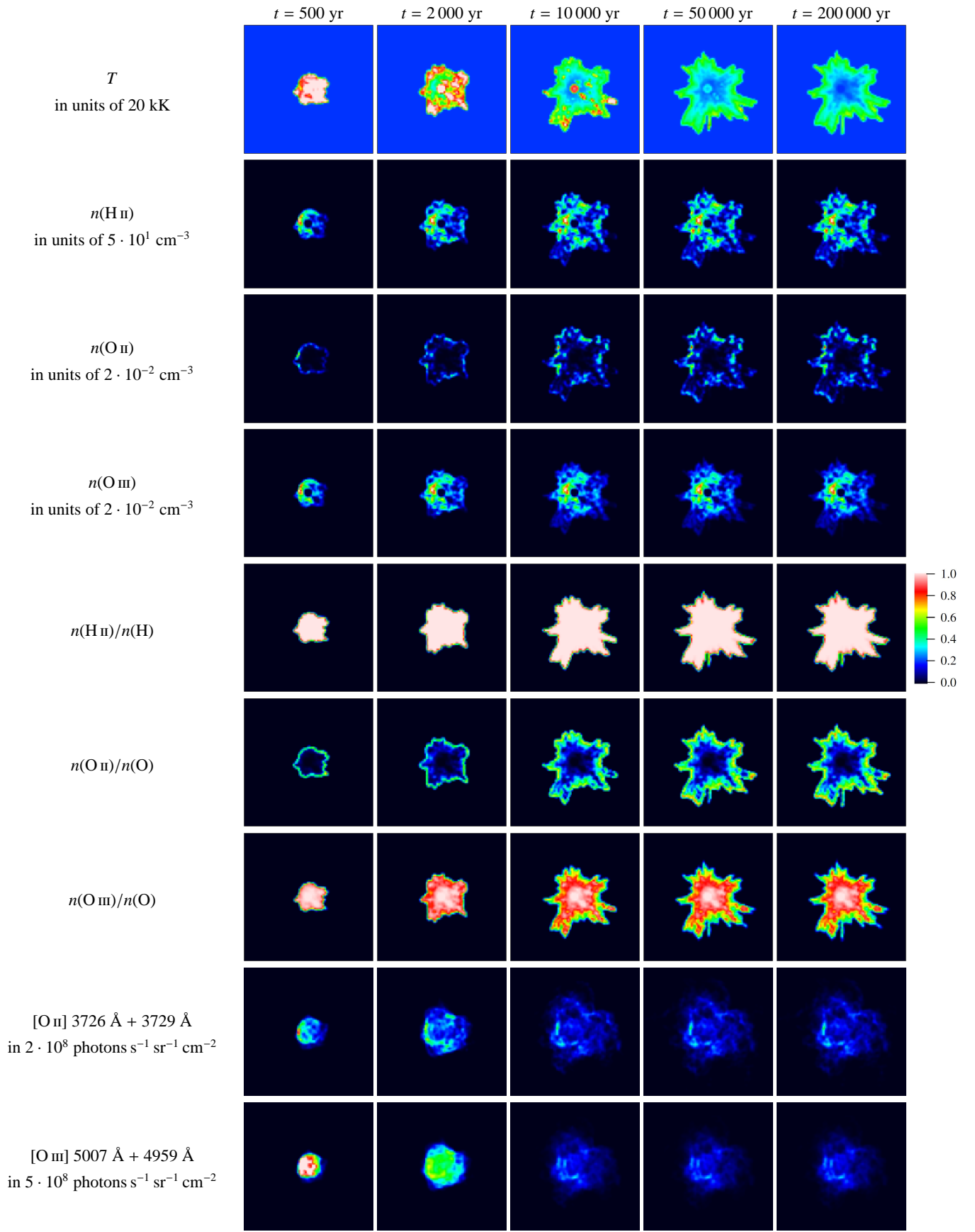

**Fig. 7.** As Fig. 6, but a solar metallicity is assumed for both the gas of the H II region and the ionizing source.

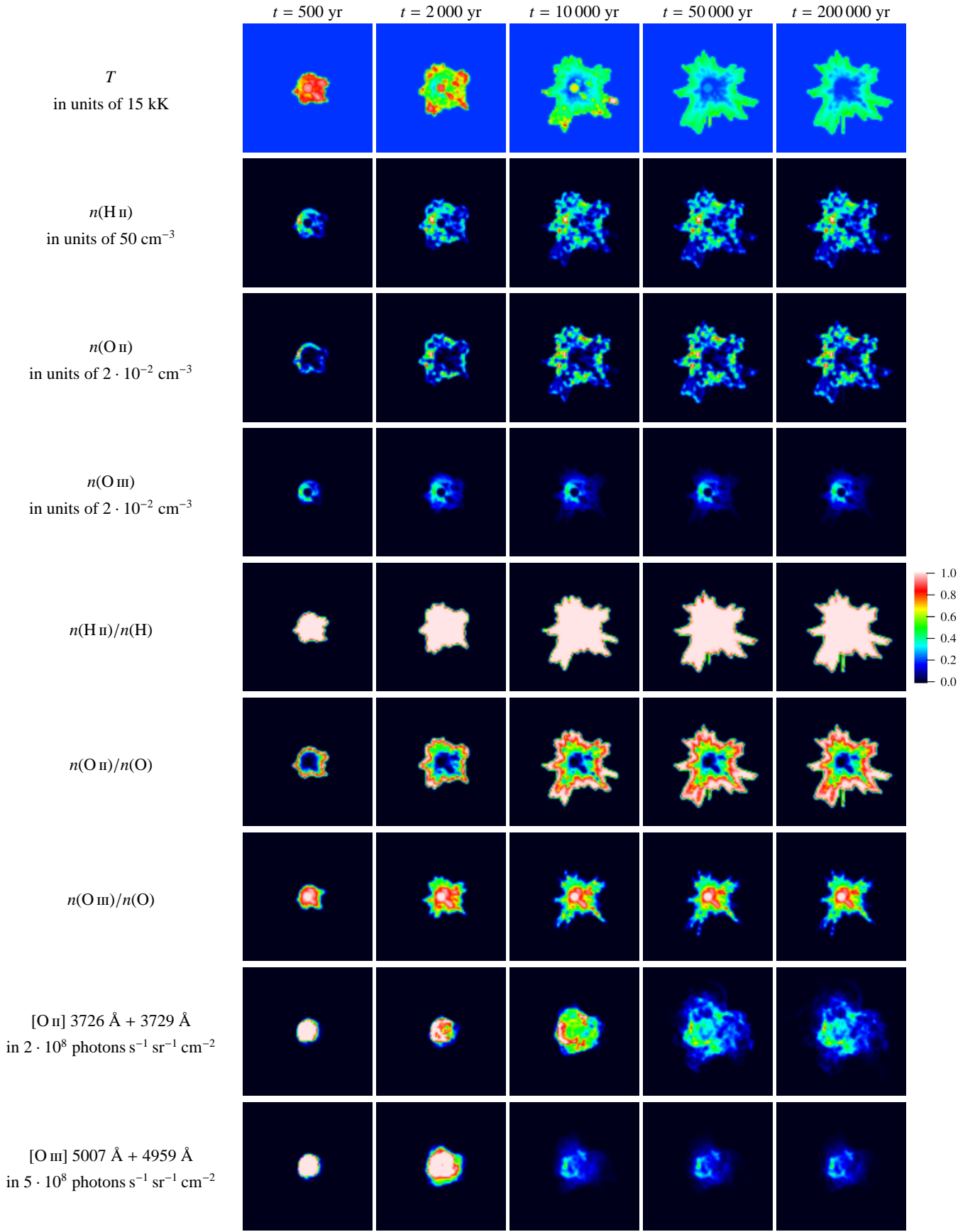

**Fig. 8.** As Fig. 7, but instead of a single 40 000 K dwarf star, two of the 35 000 K dwarf model stars and three of the 30 000 K model dwarf model stars ($1.0 Z_\odot$) are located in the center of the simulated volume. The total H-ionizing flux is 89% of flux of the 40 000 K with solar metallicity, resulting in a similar ionized volume, but due to the softer ionizing spectrum the fraction of singly ionized oxygen is larger than in the simulation with the 40 000 K star.

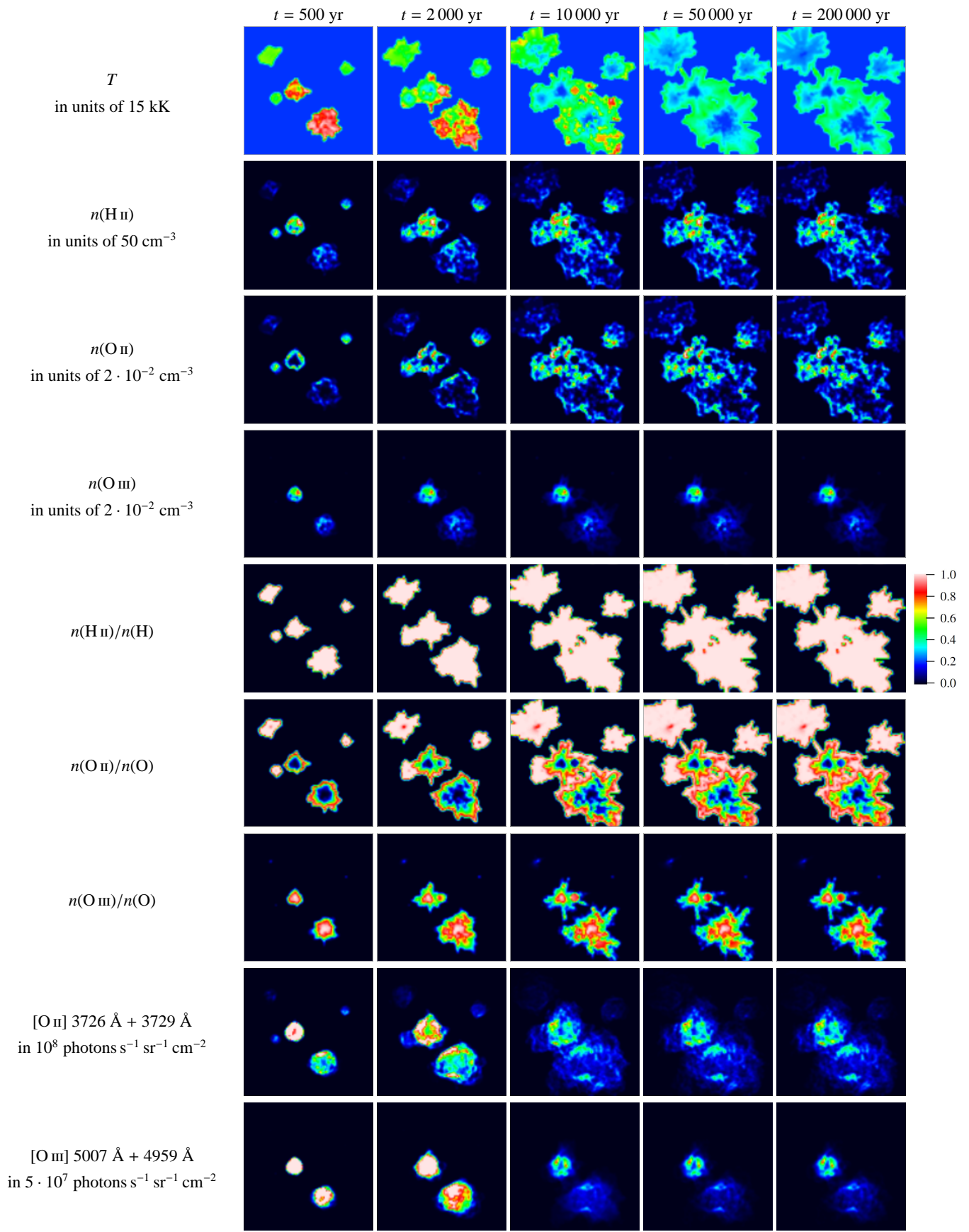

**Fig. 9.** As Fig. 8, but here the distances between the ionizing stars are similar to the extent of the H ɪɪ regions. The images show the expansion and the partial merging of the ionized volumes around these stars. The hotter (35 000 K) stars are recognized by the high temperature regions of the gas and those regions where O ɪɪɪ is the dominant ionization stage of oxygen.

**Table 7.** Comparison of ionization structures and emission properties of the homogeneous and inhomogeneous H ɪɪ regions shown in Fig. 4. In addition to the integrated ionization fractions of important elements we present the total luminosities of several observationally important lines and give for each of these lines the maximal intensity reached along the lines of sight.

| | 0.1 Z☉ | | | | | | | | 1.0 Z☉ | | | | | | | |
|---|---|---|---|---|---|---|---|---|---|---|---|---|---|---|---|---|
| | homogeneous | | | | fractal | | | | homogeneous | | | | fractal | | | |
| | \multicolumn integrated ionization fractions in % | | | | | | | | | | | | | | | |
| | I | II | III | IV | I | II | III | IV | I | II | III | IV | I | II | III | IV |
| H | 73.6 | 26.4 | | | 83.4 | 16.6 | | | 86.6 | 13.4 | | | 91.4 | 8.6 | | |
| He | 72.1 | 27.9 | 0.0 | | 82.1 | 17.9 | 0.0 | | 86.2 | 13.8 | 0.0 | | 91.3 | 8.7 | 0.0 | |
| C | 0.0 | 80.9 | 18.7 | 0.3 | 0.0 | 88.6 | 11.1 | 0.3 | 0.0 | 91.4 | 8.6 | 0.0 | 0.0 | 94.9 | 5.1 | 0.0 |
| N | 71.7 | 7.1 | 20.9 | 0.3 | 81.5 | 5.9 | 12.4 | 0.2 | 85.8 | 4.8 | 9.4 | 0.0 | 90.6 | 3.9 | 5.6 | 0.0 |
| O | 71.3 | 7.0 | 21.8 | 0.0 | 81.1 | 5.9 | 13.0 | 0.0 | 85.7 | 6.8 | 7.5 | 0.0 | 90.5 | 5.0 | 4.5 | 0.0 |
| Ne | 71.1 | 9.0 | 19.9 | 0.0 | 80.9 | 7.2 | 11.9 | 0.0 | 85.7 | 9.3 | 5.0 | 0.0 | 90.6 | 6.3 | 3.1 | 0.0 |
| S | 0.0 | 75.2 | 16.9 | 7.9 | 0.0 | 85.1 | 9.8 | 5.1 | 0.0 | 87.7 | 10.9 | 1.3 | 0.0 | 92.6 | 6.5 | 0.9 |
| | \multicolumn total line emission in units of $10^{37}$ erg s$^{-1}$ | | | | | | | | | | | | | | | |
| Hα | 1.71 | | | | 1.71 | | | | 1.80 | | | | 1.80 | | | |
| [N ɪɪ] 6584 Å + 6548 Å | 0.10 | | | | 0.11 | | | | 0.26 | | | | 0.32 | | | |
| [O ɪɪ] 3726 Å + 3729 Å | 0.37 | | | | 0.40 | | | | 0.56 | | | | 0.62 | | | |
| [O ɪɪɪ] 5007 Å + 4959 Å | 1.98 | | | | 1.84 | | | | 0.82 | | | | 0.74 | | | |
| | \multicolumn maximal line strength in units of $10^{8}$ photons s$^{-1}$ sr$^{-1}$ cm$^{-2}$ | | | | | | | | | | | | | | | |
| Hα | 0.88 | | | | 3.95 | | | | 1.88 | | | | 8.61 | | | |
| [N ɪɪ] 6584 Å + 6548 Å | 0.04 | | | | 0.26 | | | | 0.19 | | | | 1.54 | | | |
| [O ɪɪ] 3726 Å + 3729 Å | 0.09 | | | | 0.55 | | | | 0.25 | | | | 1.72 | | | |
| [O ɪɪɪ] 5007 Å + 4959 Å | 0.91 | | | | 4.07 | | | | 0.42 | | | | 2.18 | | | |

sources, and do not attempt to model a cluster SED via population synthesis.

We first assume that these stars are in close proximity to each other (which we realize by placing them in the same grid cell) and are formed instantaneously at the same time. For the composition of the cluster we assume three 30 000 K stars and two 35 000 K stars (models D-30 and D-35), which together have approximately the same (i.e., 89%) of the hydrogen-ionizing photon flux as a single of our 40 000 K dwarf star at solar metallicity, although the total luminosity of the cluster exceeds that of the 40 000 K model star by a factor of 2.8. We assume solar metallicity for both stars and gas.

The softer SED of the ionizing radiation of the cluster results in a considerably smaller fraction of N ɪɪɪ, O ɪɪɪ, Ne ɪɪɪ and S ɪᴠ as a comparison of Tables 7 and 8 shows, although the similar hydrogen-ionizing fluxes (and the identical metallicity which results in comparable temperatures and consequently recombination rates) lead to almost identical hydrogen Strömgren volumina. In the temporal evolution there is an overshooting of the temperature of the gas passed by the ionization fronts like in our simulations containing single stars. The evolution of the ionized gas around this cluster is shown in Fig. 8.

The results for the ionization fractions and also the line emission do not change significantly (i.e., by less than 15%) if the distances between the stars are increased even up to the typical sizes of the Strömgren spheres, such that no single H ɪɪ region is formed, but the ionized volumes around the stars merge only partially (see Fig. 9).

As is expected from the different SEDs, the ionization structure of oxygen around the different types of stellar sources varies significantly among the single H ɪɪ regions. In the regions around the 35 000 K stars O ɪɪ is the most abundant ionization stage of oxygen up to approximately half the distance between the sources and the border of the hydrogen Strömgren volume. Around the 30 000 K stars, the volumes in which O ɪɪɪ is more abundant than O ɪɪ are too small to be spatially resolved within our simulation, This is in agreement with the ionizing fluxes presented in Table 4 and the results of the spherically symmetric H ɪɪ region models in Sect. 3.1.2.

**Spectroscopic signature of H ɪɪ regions around very massive stars.** The direct detection of very massive stars (cf. Sect. 2.2) is problematic procedure for two reasons. First, they are rare objects such that even the closest of these object are likely to exist only on extragalactic distance scales. Second, their occurrence is correlated with massive stellar clusters where they might form by stellar mergers (cf. Pauldrach et al. 2012), such that observations of distant objects are likely to suffer from crowding effects. In this section we therefore examine the possibility of identifying VMS by means of the emission spectrum of the surrounding H ɪɪ regions. We simulate H ɪɪ regions around the 45 000 K and 65 000 K stars with $M_* = 3000\,M_\odot$, which represent the stars with the highest masses in our VMS grid, and the 40 000 K and 50 000 K stars with $M_* = 150\,M_\odot$, which are close to the currently known upper mass limit of O stars. The luminosity has been chosen to correspond to a stellar content of 3000 $M_\odot$, i.e., we simulate single 3000 $M_\odot$ stars and "clusters" of twenty 150 $M_\odot$ stars. The size of the simulation volume is (400 pc)$^3$. In the simulations all stars are located in the center this volume. In Fig. A.11 we show the emission line intensities for [O ɪɪ] 3726 Å

**Table 8.** Comparison of H ıı regions ionized by clusters with the same stellar content (two 35 000 K and three 30 000 K stars with solar metallicity), but different spatial distributions of the stars. In the first case, the distances between the stars are small when compared to the size of the Strömgren sphere. In the second case, the distances between the sources are similar to the diameters of the ionized volumes.

| | dense cluster | | | | sparse cluster | | | |
|---|---|---|---|---|---|---|---|---|
| | integrated ionization fractions in % | | | | | | | |
| | I | II | III | IV | I | II | III | IV |
| H | 91.5 | 8.5 | | | 90.0 | 10.0 | | |
| He | 95.9 | 4.1 | 0.0 | | 95.1 | 4.9 | 0.0 | |
| C | 0.0 | 97.9 | 2.1 | 0.0 | 0.0 | 97.9 | 2.3 | 0.0 |
| N | 90.9 | 7.0 | 2.1 | 0.0 | 89.1 | 8.6 | 2.3 | 0.0 |
| O | 90.9 | 8.1 | 1.0 | 0.0 | 89.2 | 9.7 | 1.1 | 0.0 |
| Ne | 91.4 | 8.4 | 0.1 | 0.0 | 90.2 | 9.7 | 0.1 | 0.0 |
| S | 0.0 | 93.7 | 6.2 | 0.1 | 0.0 | 93.4 | 6.4 | 0.1 |
| | total line emission in $10^{37}$ erg s$^{-1}$ | | | | | | | |
| H$\alpha$ | 1.80 | | | | 1.80 | | | |
| [N ıı] | 0.42 | | | | 0.44 | | | |
| [O ıı] | 0.52 | | | | 0.54 | | | |
| [O ııı] | 0.06 | | | | 0.07 | | | |
| | max. line strength in $10^8$ photons s$^{-1}$ sr$^{-1}$ cm$^{-2}$ | | | | | | | |
| H$\alpha$ | 8.49 | | | | 7.92 | | | |
| [N ıı] | 1.66 | | | | 1.30 | | | |
| [O ıı] | 1.19 | | | | 1.00 | | | |
| [O ııı] | 0.43 | | | | 0.49 | | | |

+ 3729 Å, [O ııı] 5007 Å + 4959 Å, [O ııı] 88.3 $\mu$m, and [O ıv] 25.9 $\mu$m.

The volume containing singly ionized helium is considerably smaller than the hydrogen Strömgren volume around the 3000 $M_\odot$ model star with $T_{\rm eff} = 45\,000$ K and solar metallicity, because of the strong absorption within the stellar wind of this object (see Sect. 2.2). Outside the He ı/He ıı Strömgren sphere no ionization stage of oxygen above O ıı is reached. Consequently the [O ıı] 3726 Å + 3729 Å line emission exceeds the emission of the [O ııı] 5007 Å + 4959 Å lines, which is not the case for the other dwarf or supergiant models with $T \geq 40\,000$ K. Both O ıı and O ııı exist in the part of the ionized volume where the predominant ionization stage of helium is He ıı, but the fraction of O ııı increases for higher effective temperatures of the ionizing stars. The O ıv emission line at 25.9 $\mu$m and the He ıı recombination line at $\lambda = 4696$ Å are emitted by regions where helium is ionized to He ııı, i.e., around objects with large effective temperatures such as the 65 000 K model VMS. The latter lines have been detected in starburst galaxies (Lutz et al. 1998). Here hot ($T > 50\,000$ K) very massive stars created by mergers in dense clusters may provide a complementary explanation to shocks (Lutz et al. 1998) or the photoionization by Wolf-Rayet stars (Schaerer & Stasińska 1999). The O ıv line is however strongly metallicity-dependent as for lower metallicity there is a smaller number of potential emitters and the collisional excitation rate coefficient is reduced for larger temperatures of the metal-poor gas (cf. Sect. 3.1.2). Therefore the emission of the [O ıv] by gas around very hot and very massive stars is more likely to be observed in an already chemically enriched environment.

# 4. Summary and conclusions

In this work we have presented state-of-the-art spectral energy distributions of hot, massive stars as a contribution to study of the line emission from the ionized gas around stars and stellar clusters. Such H ıı regions are essential tracers for the state and evolution of galaxies because they are closely connected to star formation processes and their emission line spectra provide information about the chemical composition, the structure, and the dynamics of the interstellar gas in galaxies, but their spectra depend to a large extent on the SEDs of the ionizing stellar sources.

Our stellar model grid comprises O-type dwarf and supergiant stars with temperatures from 30 000 K to 55 000 K and metallicities between 0.1 $Z_\odot$ and 2.0 $Z_\odot$, representing the metallicity range of star-forming regions in the present-day universe. A comparison of the ionizing fluxes for different metallicities shows that even for the same effective temperature there are considerable differences (up to several orders of magnitude) in the emission rates of photons in the energy ranges required to produce key ionic species (such as O ııı or Ne ııı, which have important diagnostic lines). In most cases the fluxes in these energy ranges decrease with increasing stellar metallicities due to the enhanced line blocking in the atmospheres of the stars.

Additionally, we have computed the ionizing fluxes of very massive stars that possibly form through collisional mergers in dense massive clusters and may considerably exceed the mass limit of O-type stars forming directly from molecular gas clouds. For these objects we have considered masses from 150 $M_\odot$ to 3000 $M_\odot$ and temperatures from 40 000 K to 65 000 K.

We have used the SEDs in spherically symmetric photoionization models to gauge the effect of stellar and gas metallicity on the ionization structure of the H ıı regions. As expected from a direct assessment of the SEDs, for a given stellar effective temperature the integrated number fractions of the different ionization stages of the metals in the gas strongly depend on the metallicity-dependent SEDs for those species whose ionization energies lie above the ionization edge of He ı. Even at the same metallicity and temperature, the results differ significantly between dwarfs and supergiants due to the influence of the much stronger stellar winds of supergiants on the SEDs.

To judge the effects of an inhomogeneous distribution of the gas on the nebular line emission we have used our 3d radiative transfer code, augmented to account for the metals C, N, O, Ne, and S, to compute integrated line-of-sight intensities and synthetic narrowband images. Although an inhomogeneous density structure leads to higher peaks in the distribution of the observable line intensities for a given mean density of the gas, we find no relative changes above 20% in the total emission per line.

With regard to the temporal evolution of the ionization and temperature structure of the H ıı regions we find an overshooting of the gas temperatures significantly above the equilibrium value during the transition of the gas from initially neutral to ionized. However, due to the short timescales, and the fact that in reality the ionizing sources themselves evolve, this is unlikely to be of practical significance for the analysis of H ıı regions with current models.

Finally, we have studied the effects of clustering of stars in two limiting cases, a dense cluster where the distances between the stars become irrelevant compared to the extent of the (combined) H ıı region, and a loose association where the distances between the stars are of the same order as the Strömgren radii, so that the individual H ıı regions only partly overlap. We find that the differences between the two models in the number frac-tions of the ionization stages and the total emission rates per line remain moderate – the total emitted fluxes of the considered emission lines differ by less than 10%.

Although spherically symmetric modeling of H ıı regions is still an important tool for determining the global properties of the ionized gas and for performing systematic examination of large sets of different input parameters, three-dimensional radiative transfer models accounting for inhomogeneous density structures, several sources of ionization, and a time-dependent evolution of the ionized medium are able to provide additional information for studying H ıı regions in more detail. A comparison of the results from these models with narrow-band or spectrally resolved observations allows obtaining detailed information about the internal structure of H ıı regions and its relation to the ionizing stars.

*Acknowledgements.* We thank an anonymous referee for helpful comments which improved the paper. We further wish to thank K. Butler for helpful discussions concerning the atomic data used for our computations and B. Ercolano for useful comments. We like to thank N. Gnedin for providing the IFrIT program used for some visualizations of our results. This work was supported by the Deutsche Forschungsgemeinschaft (DFG) under grants PA 477/18-1 and PA 477/19-1.

# Appendix A: Figures and model data

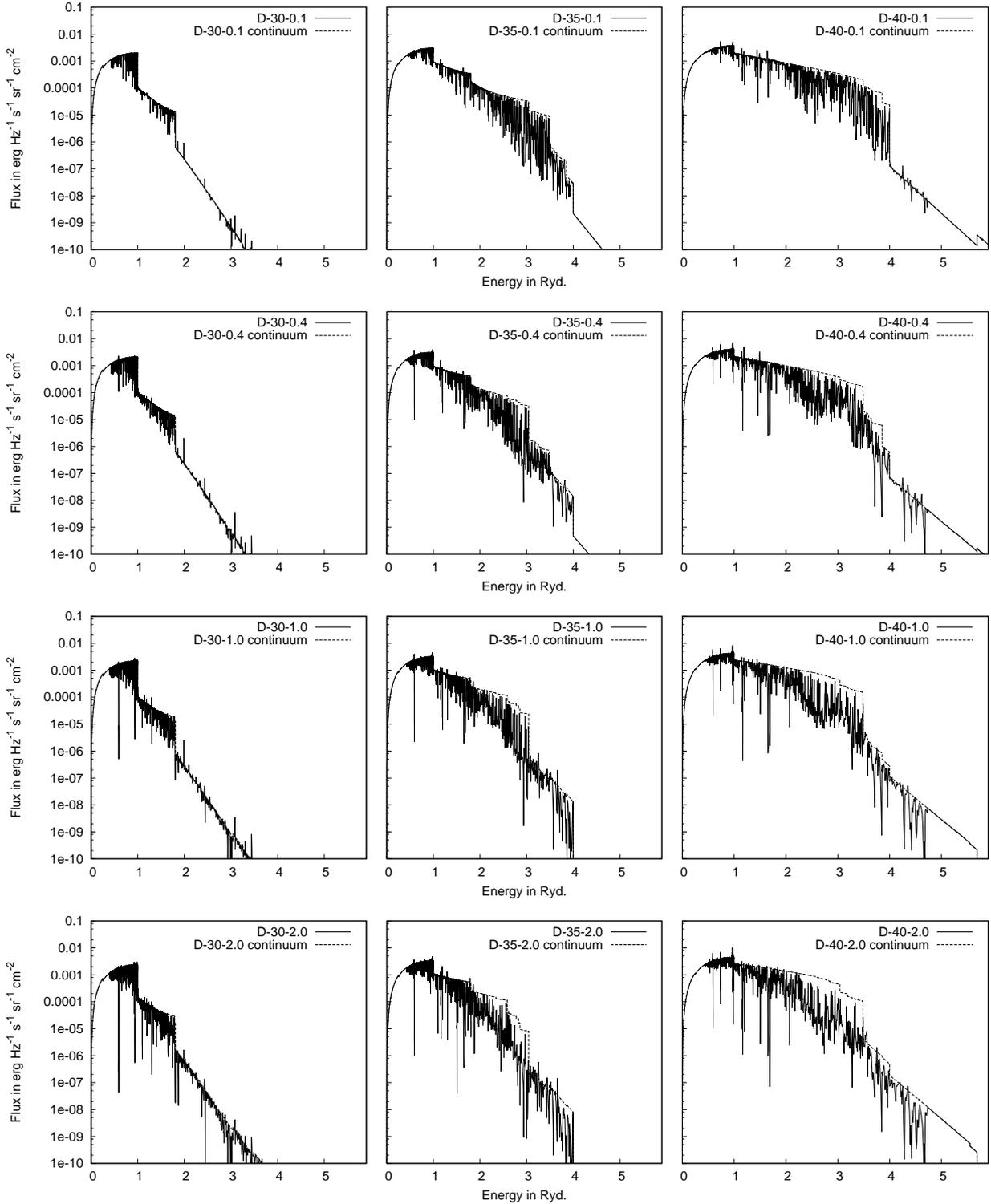

**Fig. A.1.** Spectra of the "dwarf models" for $T_{\mathrm{eff}}$ = 30 000 K (left column), $T_{\mathrm{eff}}$ = 35 000 K (intermediate column), and $T_{\mathrm{eff}}$ = 40 000 K (right column) and different metallicities (0.1, 0.4, 1.0 and 2.0 times the solar metallicity – the metallicity is increasing from the upper to the lower end of each column) – shown are the computed spectra and the corresponding continua (Eddington flux $H_\nu$ in erg/cm²/s/Hz versus the inverse wavelength in Rydberg). The data can be copied from [D30–D40 model data].[24]

---

[24] The four columns in the data file contain from left to right the wavelength in Angstroms (Å), the Eddington flux $H_\nu$ in erg/cm²/s/Hz, the continuum flux in the same units, and the normalized spectrum which is obtained by dividing the second column by third column. The flux has been normalized to correspond to the effective temperature at the photospheric radius. Note that the data may be available only on the astro-ph version of this paper – cf. ArXiv e-prints.

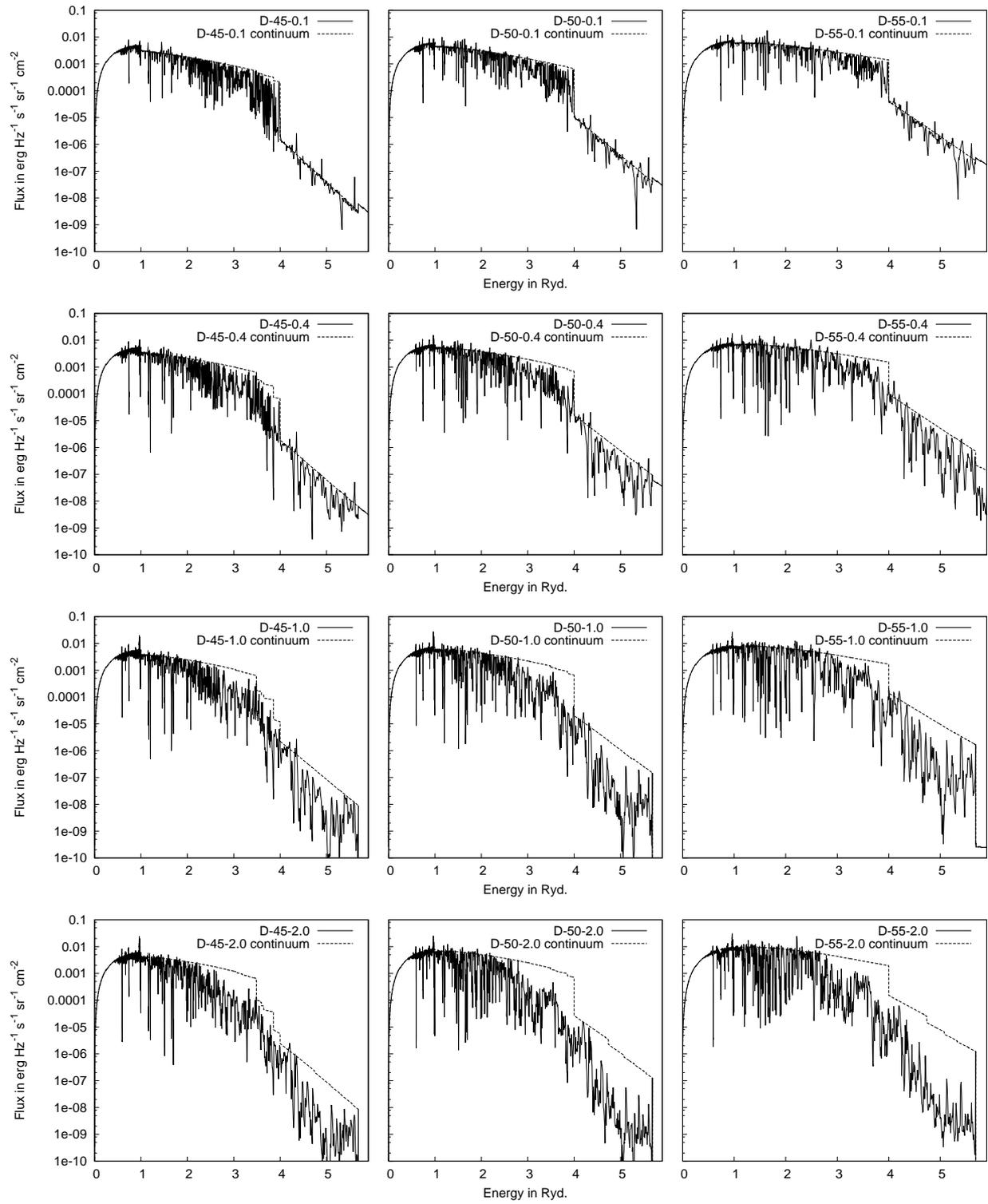

**Fig. A.2.** Same as Fig. A.1 but for $T_{\text{eff}} = 45\,000$ K, $T_{\text{eff}} = 50\,000$ K, and $T_{\text{eff}} = 55\,000$ K. The data can be copied from [D45–D55 model data].

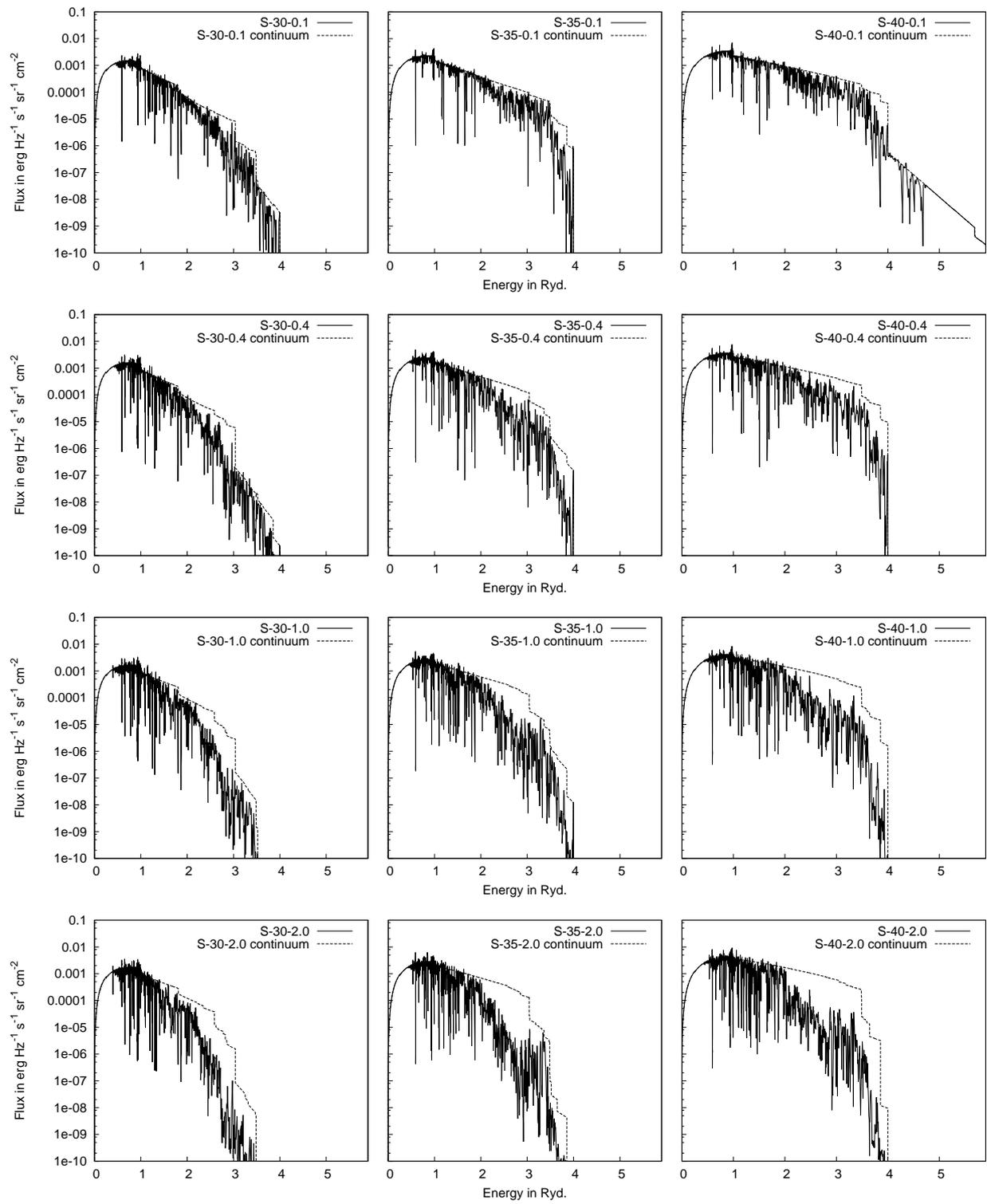

**Fig. A.3.** Same as Fig. A.1 but for "supergiant models" with $T_{\mathrm{eff}} = 30\,000$ K, $T_{\mathrm{eff}} = 35\,000$ K, and $T_{\mathrm{eff}} = 40\,000$ K. The data can be copied from [S30–S40 model data].

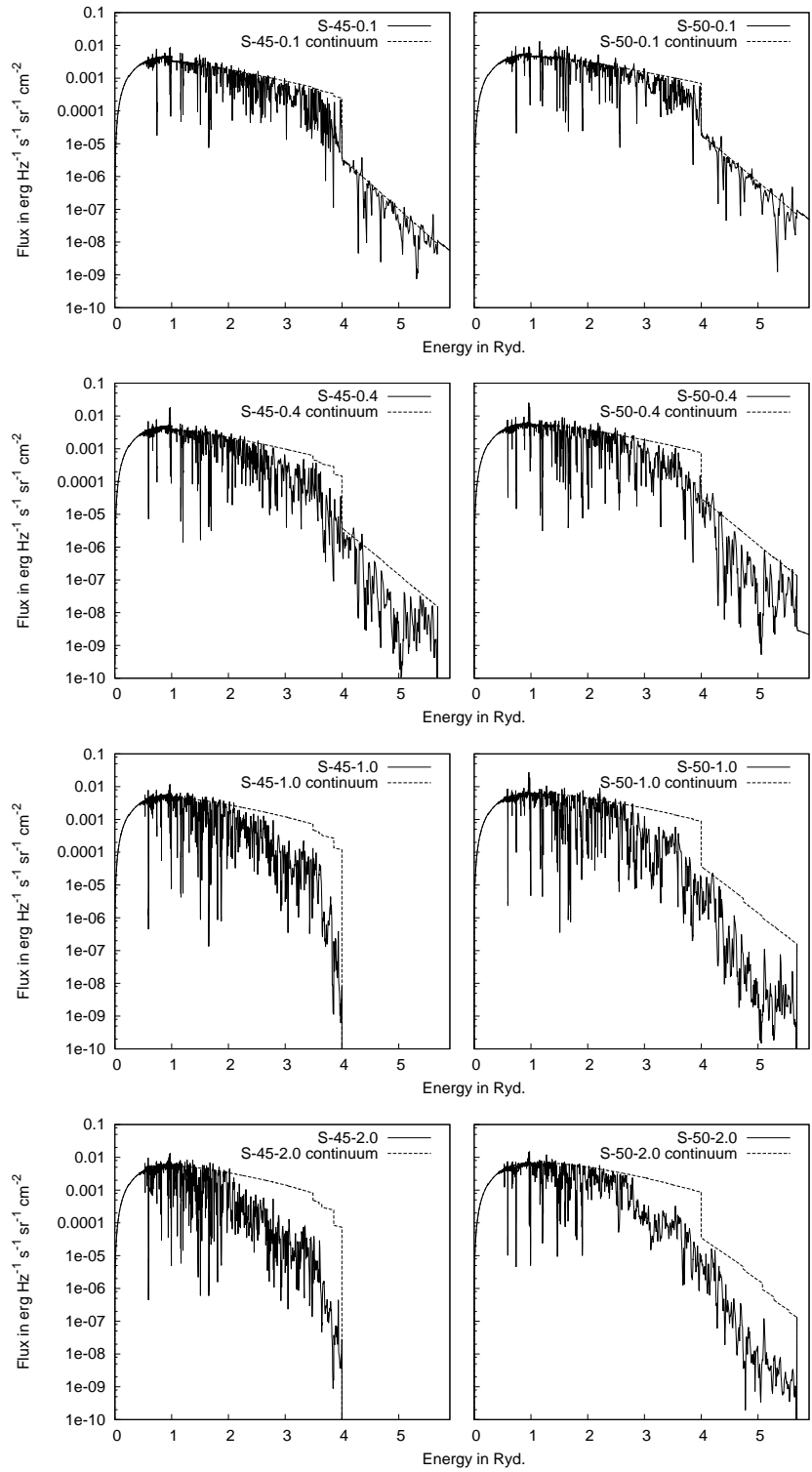

**Fig. A.4.** Same as Fig. A.1 but for "supergiant models" with $T_{\mathrm{eff}}$ = 45 000 K and $T_{\mathrm{eff}}$ = 50 000 K. The data can be copied from [S45–S50 model data].

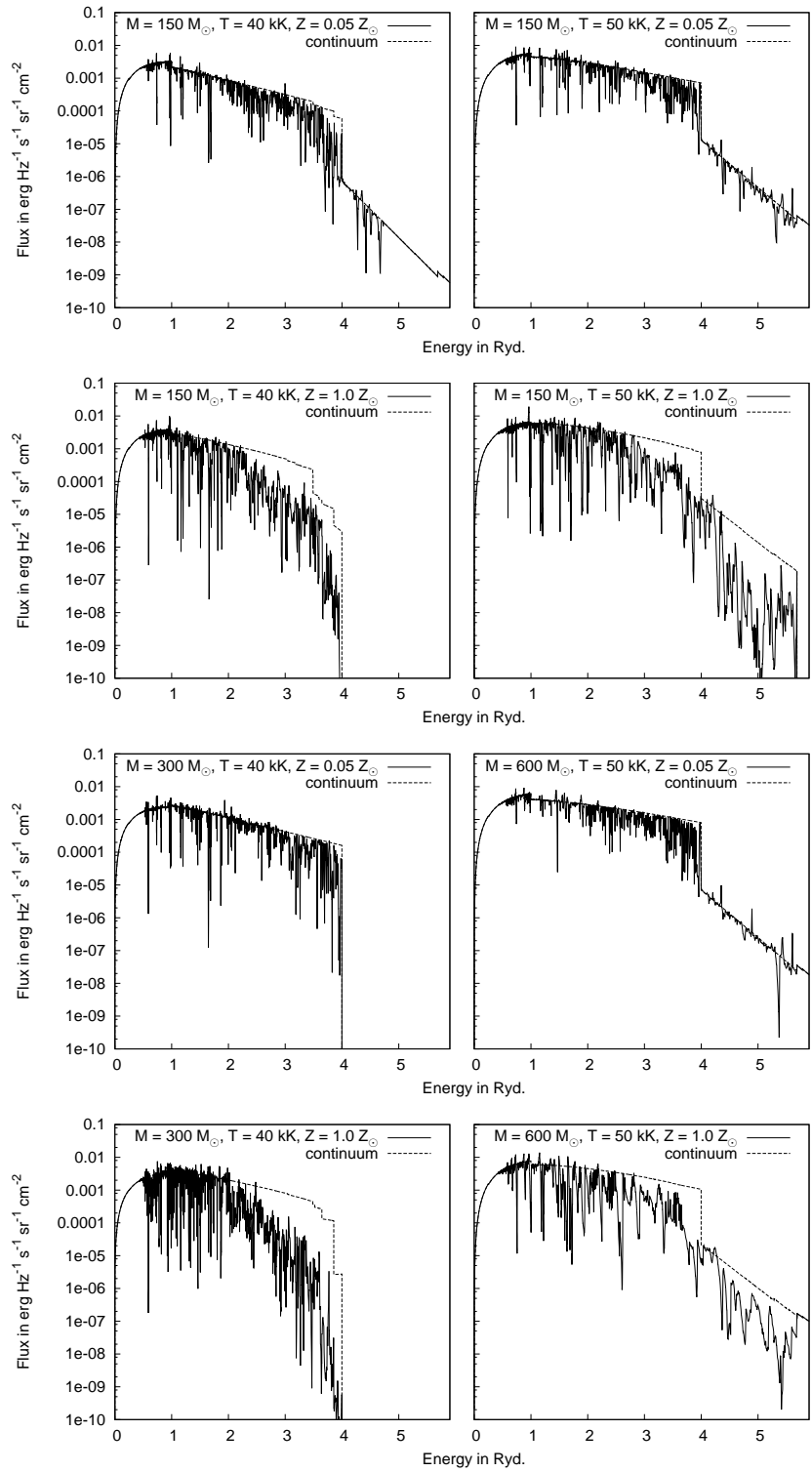

**Fig. A.5.** Same as Fig. A.1 but for "VMS models" in the mass range between $150\,M_\odot$ and $600\,M_\odot$. The data can be copied from [M150–M600 model data].

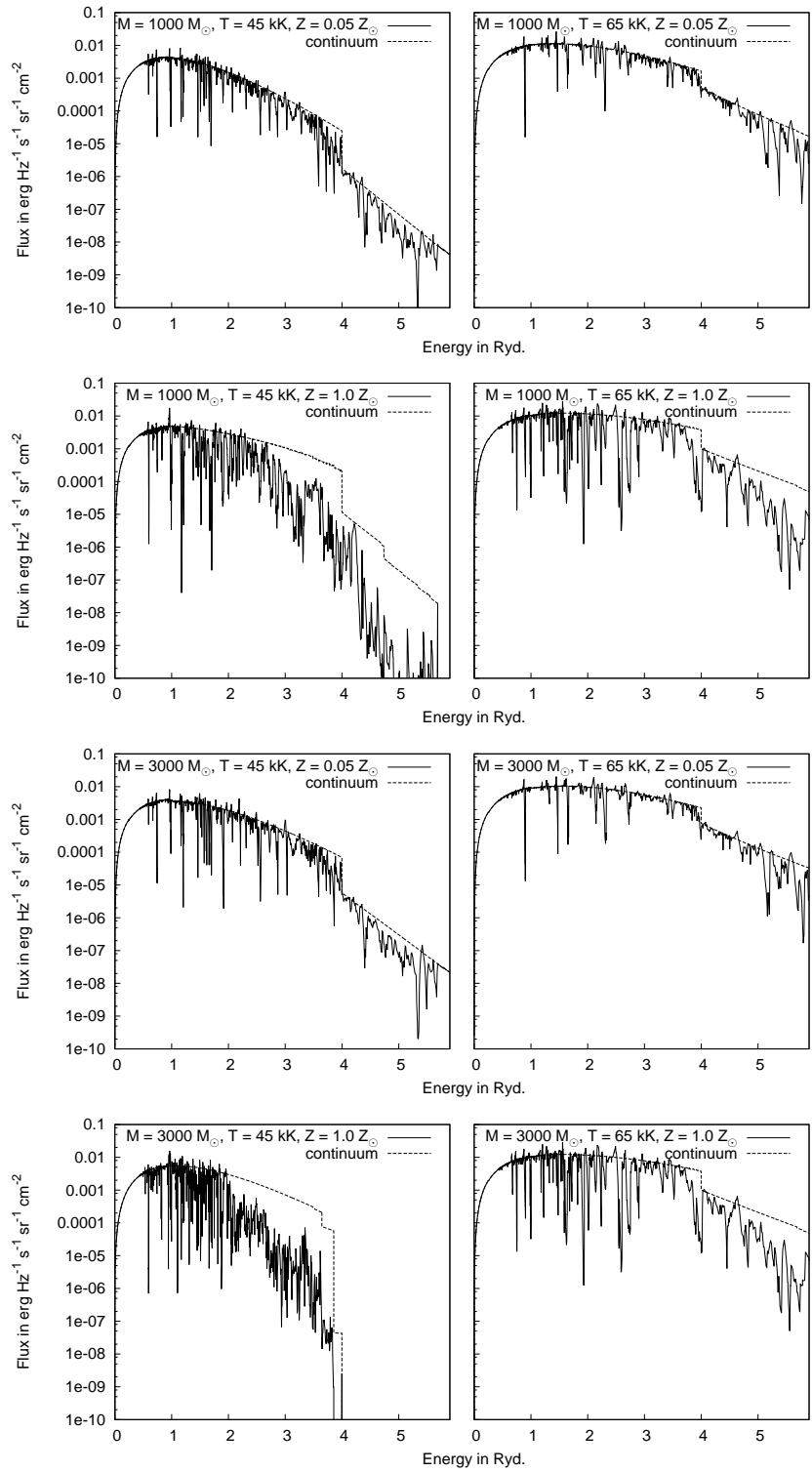

**Fig. A.6.** Same as Fig. A.1 but for "VMS models" in the mass range between 1000 $M_\odot$ and 3000 $M_\odot$. The data can be copied from [M1000–M3000 model data].

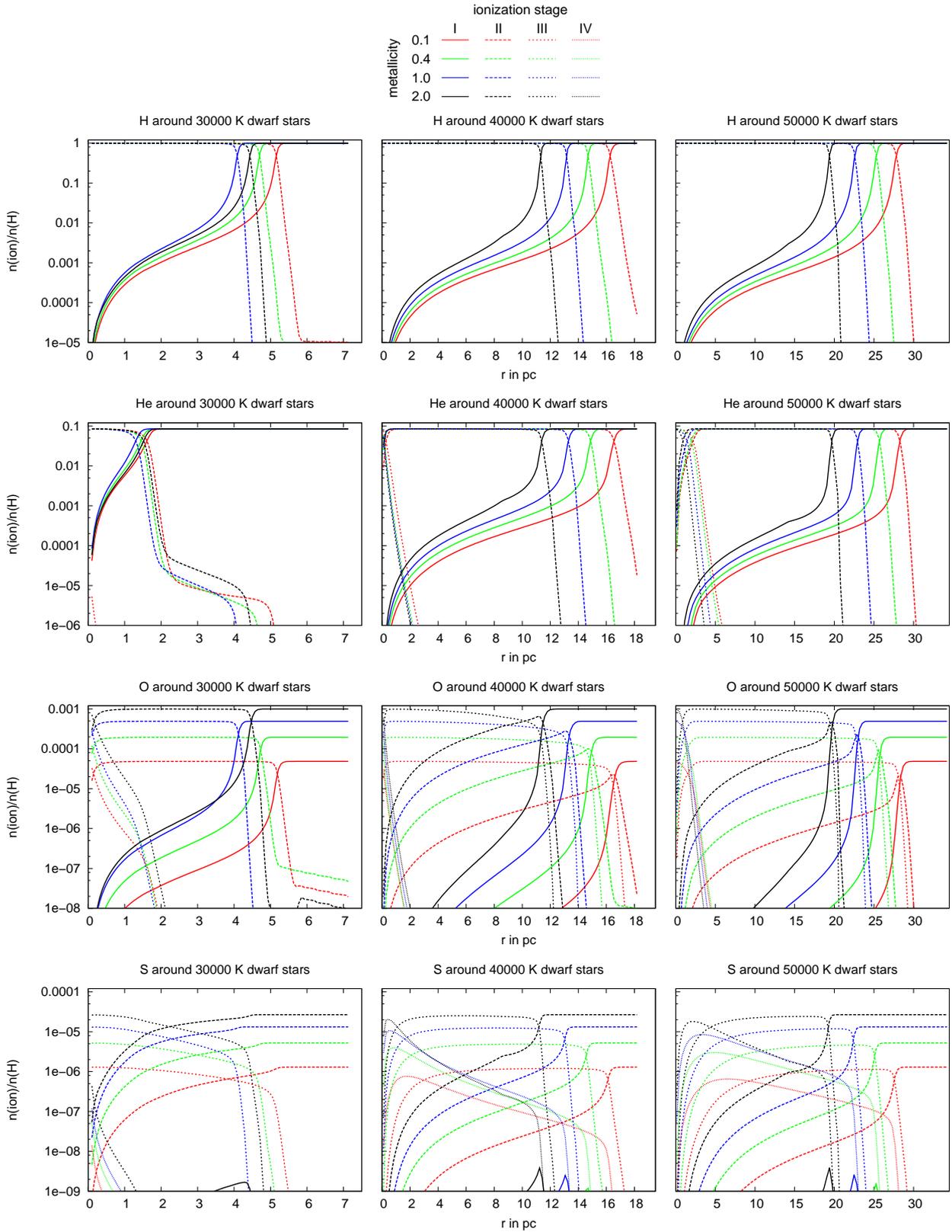

**Fig. A.7.** Ionization structures of hydrogen, helium, oxygen and sulfur of H II regions around dwarf stars of effective temperatures of 30 000 K, 40 000 K, and 50 000 K at certain values of metallicity (a homogeneous gas with a hydrogen particle density of 10 cm$^{-3}$ and the same chemical composition as the central star is assumed). As the number of hydrogen-ionizing photons varies only slightly for stars with the same effective temperature and luminosity class (cf. Table 4), the larger hydrogen Strömgren radii obtained for lower metallicities are primarily caused by the larger temperatures of the gas which in turn are caused mainly by the direct influence of the lower metallicities of the gas (see Fig.A.9).

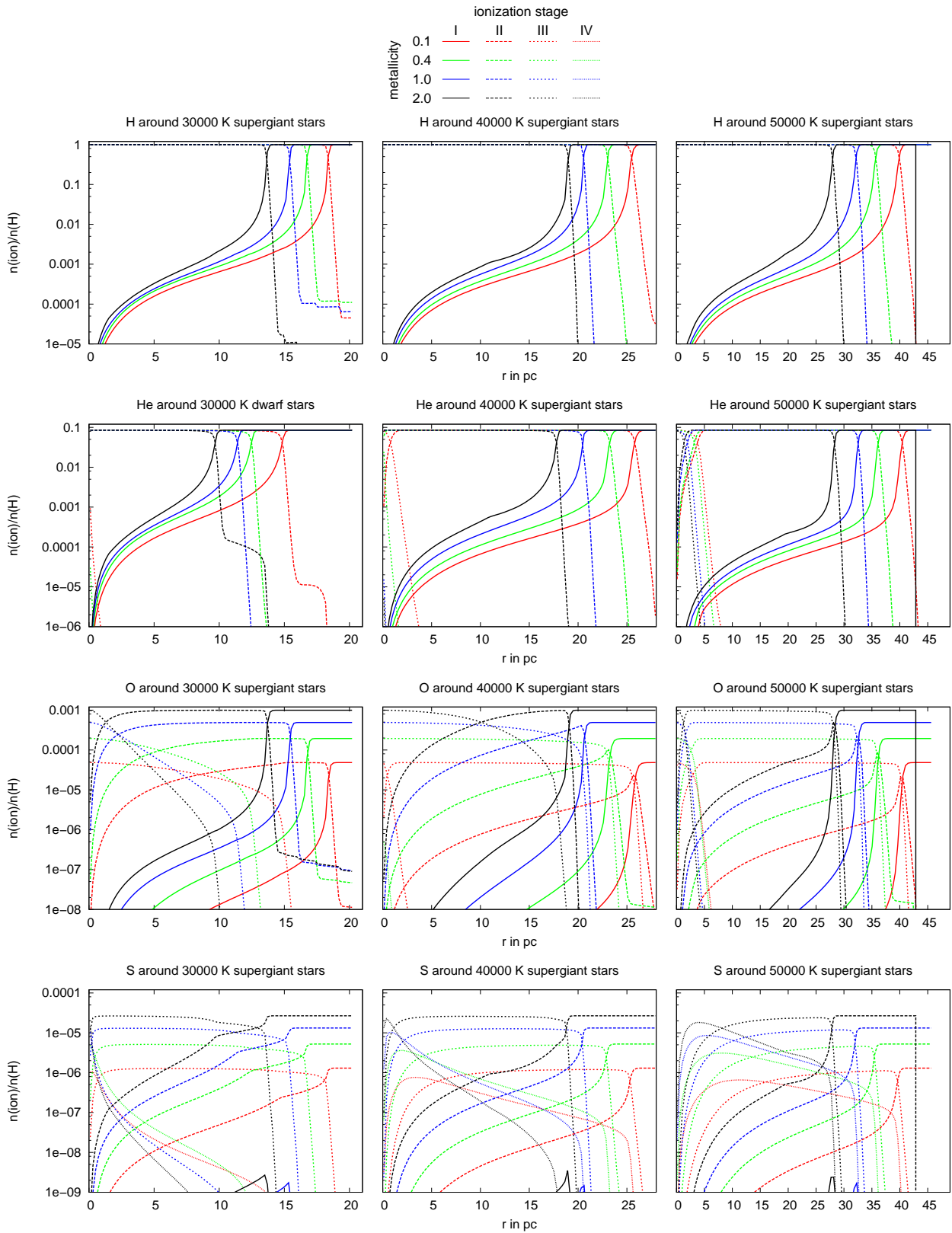

**Fig. A.8.** As Fig. A.7, but using supergiants as ionizing sources.

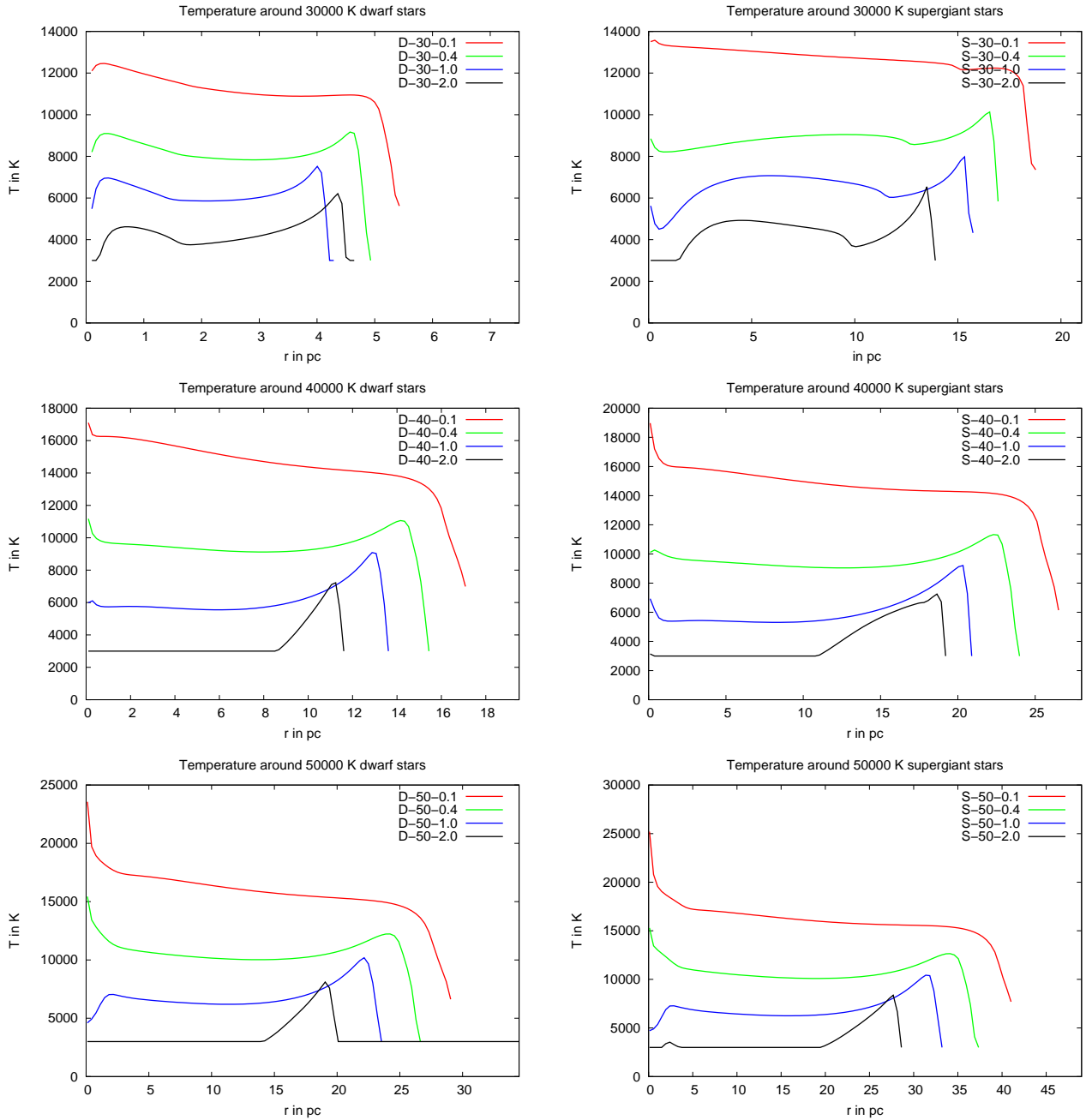

**Fig. A.9.** Temperature structures of H II regions around dwarf stars and supergiants of effective temperatures of 30 000 K, 40 000 K, and 50 000 K at certain values of metallicity (using the same physical conditions for the gas as in Fig. A.7). The influence of the metallicity on the temperature of the gas is primarily based on the reduced number of metal particles that contribute to the cooling of the gas – metal-poorer H II regions are therefore characterized by considerably higher temperatures than H II regions with larger metal abundances –, but it is also based on the harder ionization spectra obtained for metal-poorer stars (see Figs. A.1–A.4) – this results in a larger energy input by photoionization heating and therefore the temperature of the gas rises with decreasing metallicity of the irradiating stars (see also Fig. A.10).

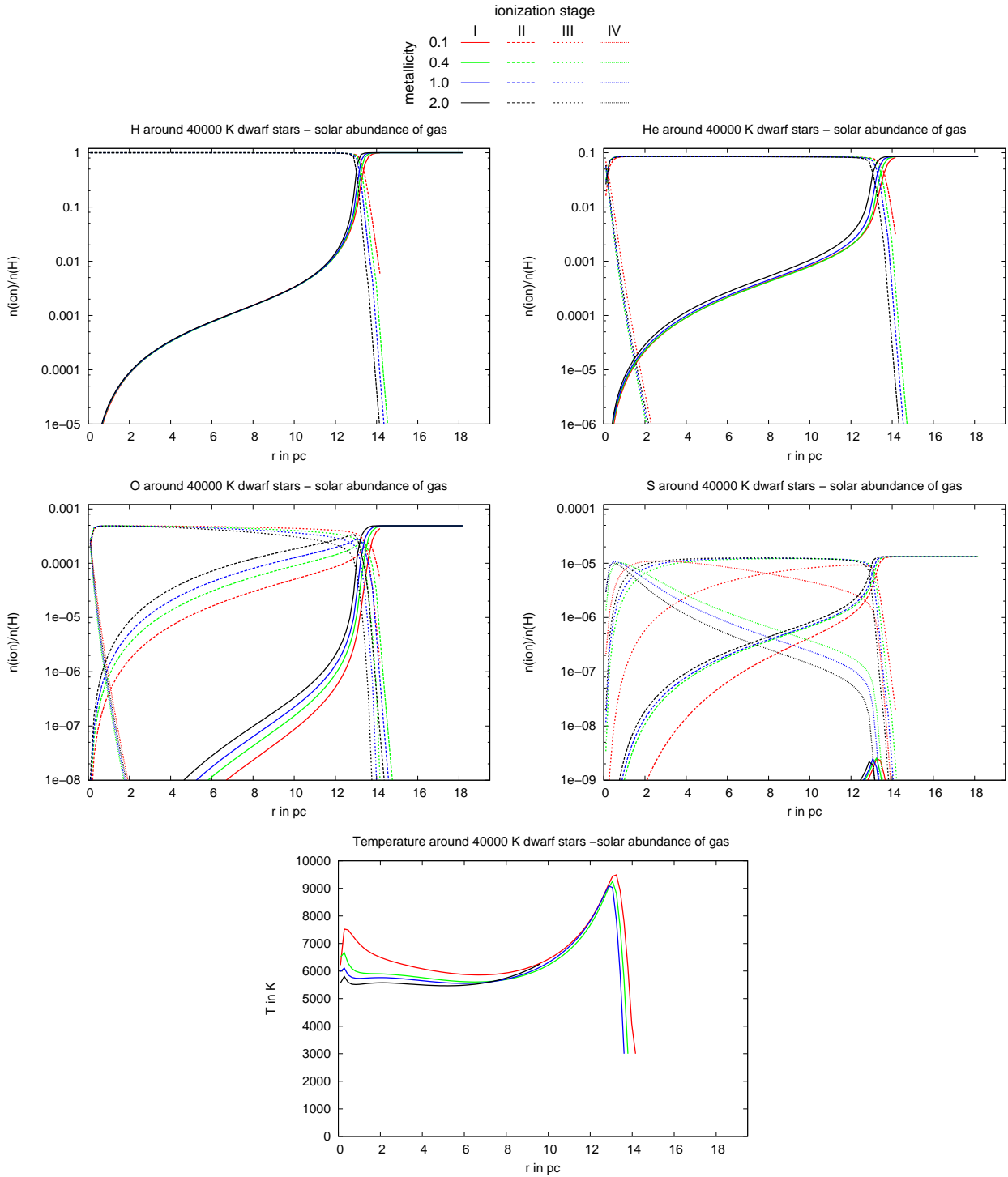

**Fig. A.10.** Comparison of the ionization structures of H, He, O, and S, and the temperature structure of H II regions for cases where the chemical composition of the gas is not the same as that of the irradiating star (for the gas a hydrogen particle density of 10 cm$^{-3}$ and solar metallicity is assumed, whereas for the stars a 40 000 K dwarf model with metallicity from 0.1 $Z_\odot$ to 2.0 $Z_\odot$ is used). In this way the influence of the metallicity-dependent ionizing spectra on the physical behavior of the gas can be investigated independently of the influence of the metallicity of the gas. As the stellar spectra differ for different metallicities especially in the energy range above the ionization edge of He I, the differences obtained for the ionization structures of O and S are more pronounced than those of H and He. And these differences lead to enhanced ionization fractions of O II and S III for the spectra of low-metallicity stars, because the spectra of these stars are "harder" than those of high-metallicity stars of the same effective temperature (see Figs. A.1–A.4). As a consequence of this behavior the temperature of the gas increases also, but this effect is smaller than in the case where the metallicity of the gas is the same as that of the star (cf. Fig. A.9).

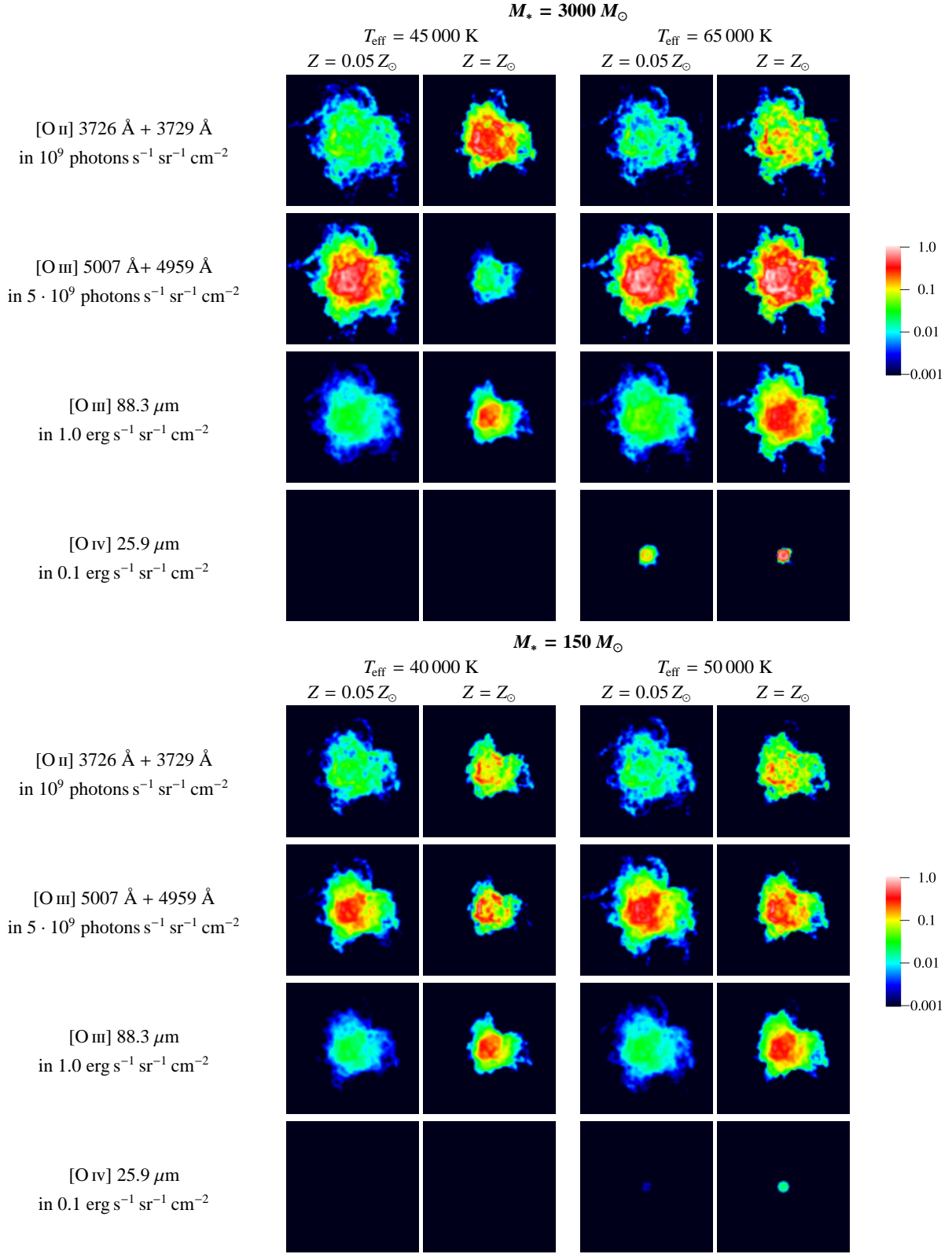

**Fig. A.11.** Synthetic images showing the intensities of important emission lines connected to various ionization stages of oxygen around very massive stars with metallicities of 0.05 Z$_\odot$ and 1.0 Z$_\odot$. Concerning the optical emission lines, the lower abundances of metal ions are partly compensated by the higher temperatures and the therefore higher rate coefficients for collisional excitation. The values are shown using a logarithmic color scale as the emission line strengths vary strongly between the different models.